\documentclass[useAMS,twocolumn,usenatbib]{mn2e}
\usepackage{amsmath}
\usepackage{rotating, graphicx}
\usepackage{hyperref}
\usepackage{color}
\usepackage{subfigure}

\newcommand{\ahf}{\textsc{AHF}}
\newcommand{\subfind}{\textsc{SUBFIND}}

\newcommand{\gadget}{\textsc{gadget-3}}
\newcommand{\gadgettwo}{\textsc{gadget-2}}

\newcommand{\rockstar}{\textsc{ROCKSTAR}}

\newcommand{\stf}{\textsc{STF}}

\def\be{\begin{equation}} 
\def\en{\end{equation}}

\def\vmax{$v_{\rm max}$}

\renewcommand{\S}{Section~}
\newcommand{\Aq}[1]{\texttt{Aq-#1}}

\newcommand{\revision}[1]{\textcolor{black}{#1}}

\begin{document}

%----------------------------- title  -----------------------------
\title[Subhaloes gone Notts: Subhaloes as tracers of the dark matter halo shape]
{Subhaloes gone Notts: Subhaloes as tracers of the dark matter halo shape}

%--------------- authors--------------- 
\author[Hoffmann et al. ]
{\parbox{\textwidth}
{
Kai Hoffmann\thanks{E-mail: hoffmann@ice.cat}$^{1}$, 
Susana Planelles$^{2,3}$,
Enrique Gazta\~{n}aga$^{1}$, 
Alexander Knebe$^{4}$,
Frazer R. Pearce$^{5}$,
Hanni Lux$^{5,6}$,
Julian~Onions$^{5}$,
Stuart~I.~Muldrew$^{5}$,
Pascal Elahi$^{7,8}$,
Peter Behroozi$^{9}$,
Yago Ascasibar$^{4}$,
Jiaxin Han$^{7,10}$,
Michal Maciejewski$^{11}$,
Manuel E. Merchan$^{12}$,
Mark Neyrinck$^{13}$,
Andr\'{e}s N. Ruiz$^{12}$,
Mario A. Sgro$^{12}$
}
%--------------- institutes--------------- 
\vspace{0.4cm} \\
$^{1}$Institut de Ci\`{e}ncies de l'Espai (ICE, IEEC/CSIC), E-08193 Bellaterra (Barcelona), Spain\\
$^{2}$Astronomy Unit, Department of Physics, University of Trieste, via Tiepolo 11, I-34131 Trieste, Italy\\
$^{3}$INAF, Osservatorio Astronomico di Trieste, via Tiepolo 11, I-34131 Trieste, Italy\\  
$^{4}$Departamento de F\'isica Te\'{o}rica, M\'{o}dulo 15, Facultad de Ciencias, Universidad Aut\'{o}noma de Madrid, 28049 Madrid, Spain\\
$^{5}$School of Physics \& Astronomy, University of Nottingham, Nottingham, NG7 2RD, UK\\
$^{6}$Department of Physics, University of Oxford, Denys Wilkinson Building, Keble Road, Oxford, OX1 3RH, UK\\
$^{7}$Key Laboratory for Research in Galaxies and Cosmology, Shanghai Astronomical Observatory, Shanghai 200030, China\\
$^{8}$Institute for Astronomy, The University of Sydney, NSW 2006, Australia\\
$^{9}$Space Telescope Science Institute, Baltimore, MD 21218, USA\\
$^{10}$Institute for Computational Cosmology, Department of Physics, Durham University, South Road, Durham DH1 3LE, UK\\
$^{11}$Max-Planck-Institut f\"ur Astrophysik, Garching, Karl-Schwarzschild-Stra§e 1, 85741 Garching bei Muenchen, Germany\\
$^{12}$Instituto de Astronom\'ia Te\'{o}rica y Experimental (CCT Cordoba, CONICET, UNC), Laprida 922, X5000BGT, Cordoba, Argentina\\
$^{13}$Department of Physics and Astronomy, The Johns Hopkins University, 3400 N. Charles St., Baltimore, MD 21218, USA\\
}

\date{Received date / Accepted date}

\maketitle

%----------------------------- abstract  -----------------------------
\begin{abstract}
We study the shapes of subhalo distributions from four dark-matter-only simulations of Milky Way type haloes.
Comparing the shapes derived from the subhalo distributions at high resolution to those of the underlying dark matter fields we find the former to be more triaxial if the
analysis is restricted to massive subhaloes. For three of the four analysed haloes the increased triaxiality of the distributions of massive subhaloes
can be explained by a systematic effect caused by the low number of objects. Subhaloes of the fourth halo show indications
for anisotropic accretion via their strong triaxial distribution and orbit alignment with respect to the dark matter field.
\revision{These results are independent of the employed subhalo finder.}
Comparing the shape of the observed Milky Way satellite distribution to those of high-resolution subhalo samples from simulations, we find
an agreement for samples of bright satellites, but significant deviations if faint satellites are included in the analysis. These deviations
might result from observational incompleteness.
\end{abstract}

 %-----------------------------  key words  -----------------------------

%----------------------------- sections  -----------------------------
%%%%%%%%%%%%%%%%%%%%%%%%%%%%%%%%%%%%%%%%%%%%%%%%%%%%%%%%%%%%%%%%%%%%%%%%%%%%%%%%%%%%%%%%%%%
%%%%%%%%%%%%%%%%%%%%%%%%%%%%%%%%%%%%%%%%%%%%%%%%%%%%%%%%%%%%%%%%%%%%%%%%%%%%%%%%%%%%%%%%%%%
\section{Introduction}
\label{sec:intro}

The modelling of how galaxies trace the underlying matter field is one of the biggest uncertainties in observational cosmology.
Understanding the relation between the distributions of  galaxies and matter at small scales, such as those of galaxy groups and clusters,
turns out to be difficult due to the non-linear evolution of matter density fluctuations and the complex processes involved in galaxy formation.
However, it is worth addressing this challenge as modelling the  galaxy distribution at these scales increases the statistical
power of cosmological surveys. Furthermore, it opens the possibility to compare model predictions to the well observed properties of our local Universe.

The host haloes of galaxy groups and clusters can be characterised by their radial density profiles and their triaxiality.
In this study we are interested in the latter property as we analyse the relation between the shape of dark matter haloes
obtained from high resolution simulations and their subhalo populations.
The relation between these shapes is relevant for the Halo Occupation Distribution models,
which  are used to predict and interpret observational data.  In these models, satellite galaxies are distributed in host haloes with an assumed ellipticity. This
ellipticity has an impact on the correlation functions derived from these models \citep[e.g.][]{Smith_2005,  Smith_2006, vanDaalen_2012}.

Observations of the local Universe suggest that the shapes of host haloes and those of their satellite populations
are not necessarily the same since the satellites are found to be distributed along thin discs.
During recent years, the discs of satellite galaxies around the Milky Way \citep[MW;][]{lynden_bell_1976, kunkel_demers_1976}
and the Andromeda galaxies \citep[M31;][]{koch_grebel_2006, metz_2007, metz_2009, Pawlowski12, conn13} are
discussed in the literature with focus on whether or not such flat satellite distributions around MW-like host haloes are expected in
the standard, spatially flat  {\it $\Lambda$-Cold Dark Matter}  model  with cosmological constant \citep[$\Lambda$CDM,][]{Blumenthal_1984}.

\revision{Within this hierarchical model of structure formation,
haloes accrete matter from preferential directions of the surrounding sheets and filaments. As a 
consequence, the substructure distribution in haloes is not expected to be spherical
until the orbits of the accreted matter components are completely randomized \citep{Knebe_04, Zentner_2005, Libeskind_2005}.
Such an anisotropic accretion of substructures was first observed in simulations by \citet{Tormen_1997} 
\citep[see][for recent studies]{VeraCiro_2011, Lovell11, Libeskind_2012}.}
Indeed, flat satellite systems are also found in CDM N-body simulations, albeit those are unlikely to be as flat as the  satellite systems of the MW and M31
\citep[see e.g.][who also take into account possible obscuration effects from the galactic disc]{Wang_2013}. However, $\Lambda$CDM
simulations of the local Universe with constrained initial conditions are able to reproduce the galaxy distributions in our neighbourhood.
Analysing a set of such simulations, \citet{libeskind11} found that the local satellite distribution probably results from anisotropic accretion of matter.

\revision{Light was shed from a different angle on the anisotropic accretion scenario by \citet{Danowich_12}. Simulating dark matter together with gas
in hydrodynamical simulations these authors find that, at high redshifts, a large fraction of the baryonic mass of massive haloes is accreted from a few
narrow streams of cold gas which tend to lie in the same sheets of the dark matter distribution. This effect leads to thin discs of
satellites \citep{Goerdt13} and might explain why planar subhalo distributions as thin as the MW and M31 are unlikely to be found
in DM--only simulations. }

\revision{However, discrepancies between the discs of satellites in $\Lambda CDM$ simulations and observations of the local group as well as the missing satellite
problem \citep[e.g.][]{Weinmann_2006} motivate the exploration of alternative formation scenarios, within which the discs of satellites could originate
from tidal arms developed during an encounter of a young MW with another galaxy \citep[e.g.][suggested M31 for the large magellanic cloud]{Pawlowski12}. Such events are observed
\citep[e.g.][]{Weilbacher_2003}  and can be also seen in simulations \citep[e.g.][]{Bournaud_2008}
without requiring the assumption of cold dark matter.}

In this study we aim to improve the understanding of why massive subhaloes in $\Lambda$CDM simulations are distributed anisotropically
around the centre of their host halo. We therefore examine the shape of the dark matter subhalo populations in four MW-like host haloes from the
Aquarius simulation to disentangle systematic effects from anisotropic accretion as possible explanation for the flat distribution of the most massive subhaloes.
Two possible systematics are studied:

i) Substructures in dark matter haloes are not well defined objects. After accretion, dark matter particles can move away from the subhalo,
as the latter orbits under the action of the complex gravitational field of its host. The question if a given dark matter particle is still associated
to the subhalo within which it entered the host can be answered differently by different subhalo finders. The resulting uncertainty of subhalo properties
(e.g. mass) causes the problem that subhalo samples, defined by these properties, can consist of different objects, depending on the employed subhalo finder.
Uncertainties of subhalo properties can therefore propagate into uncertainties in the shape of their distribution if, for instance, the $N$ most massive subhaloes
are analised.

ii) Besides the detection of subhaloes in simulations, the small number of massive (observationally speaking, the most luminous) subhaloes complicates the
shape measurement of their distribution. Ensembles with low numbers of objects are more likely to appear aligned, which can cause an apparent, non-physical flattening of
subhalo distributions.

Looking for indications of anisotropic infall, we also investigate the orbit orientation of subhaloes, given by the direction of their angular momenta,
with respect to the dark matter field of the corresponding host haloes.

Finally, after investigating possible reasons for the flatness of the Aquarius subhalo populations,
we compare the shape of these populations to results computed from the observed MW satellite distribution.

This paper is organised as follows: 
in Section~\ref{sec:Data} we describe the simulations
analysed in this work together with the different subhalo finders participating in this study;
in Section~\ref{sec:shape_measurement} we introduce the method used to derive halo shapes.
Shape measurements of haloes and their subhalo population are presented in Section~\ref{sec:results} together
with an analysis on subhalo orbit orientation with respect to the dark matter field and a comparison
with shapes of the observed MW satellite distribution.
Finally, our findings are summarised and discussed in Section \ref{sec:summary}.
In the Appendix, we study the accuracy and precision of our shape measurements and present the observational data.
%%%%%%%%%%%%%%%%%%%%%%%%%%%%%%%%%%%%%%%%%%%%%%%%%%%%%%%%%%%%%%%%%%%%%%%%%%%%%%%%%%%%%%%%%%%
%%%%%%%%%%%%%%%%%%%%%%%%%%%%%%%%%%%%%%%%%%%%%%%%%%%%%%%%%%%%%%%%%%%%%%%%%%%%%%%%%%%%%%%%%%%
\section{Simulations}\label{sec:Data}

We study subhalo populations from four simulated Milky Way sized host haloes.
These haloes are part of the Aquarius simulation project \citep{Springel_2008}
- a suite of high resolution $N$-body simulations performed using the cosmological
SPH code \gadget\ \citep[based on \gadgettwo,][]{Springel_2005a}.
The simulated haloes were selected from the larger cosmological $\Lambda$CDM simulation Millennium-II
\citep{Boylan-Kolchin_2009} and then re-simulated at higher resolution with the cosmological parameters
$\Omega_m=0.25$, $\Omega_{\Lambda} = 0.75$, $\sigma_8=0.9$, $n_s=1$
and $H_0=100 h$ km s\textsuperscript{-1}Mpc\textsuperscript{-1} $ =
73$ km s\textsuperscript{-1}Mpc\textsuperscript{-1}. 

The four Aquarius haloes analysed in this work, labeled from \Aq{A} to \Aq{D}, have final masses of
$\sim 10^{12}$M$_\odot$ and were chosen to be relatively isolated at redshift $z=0$. Each halo
has been re--simulated at five resolution levels, labeled as 5, for the lowest, to 1,
for the highest resolution level, respectively. The mass per particle varies from
$m_p = 2.94\times 10^6 h^{-1}$M$_\odot$ in level $5$ to $m_p = 1.25\times 10^3 h^{-1}$M$_\odot$
in the highest resolution run, resolving a given halo with approximately half-million up to 1.5-billion
particles within the virial radius for level $5$ and $1$, respectively.
\revision{In our analysis we focus on subhaloes detected  at the second highest  level of resolution $2$ at $z=0$.
Those are compared to the dark matter field at the second lowest  level of resolution 4 (dark matter data from
higher resolution runs was not available for this analysis). The particle masses of these haloes are shown
in Table~\ref{table:haloes}. More information
on the haloes is given by \citet{Springel_2008}.}
\begin{table}
\centering
\begin{tabular}{cc}
\hline
Halo & particle mass [$h^{-1}{\rm M}_\odot]$\\
\hline
%Aq-A-1   &  $2.345\times 10^3$ \\
Aq-A-2   &  $1.877\times 10^4$ \\
%Aq-A-3   &  $6.727\times 10^4$ \\
Aq-A-4   &  $5.382\times 10^5$ \\
%Aq-A-5   &  $4.305\times 10^6$ \\
\hline
Aq-B-2   &  $8.832\times 10^3$ \\
Aq-B-4   &  $3.071\times 10^5$ \\
\hline
Aq-C-2   &  $1.916\times 10^4$ \\
Aq-C-4   &  $4.401\times 10^5$ \\
\hline
Aq-D-2   &  $1.914\times 10^4$ \\
Aq-D-4   &  $3.667\times 10^5$ \\
\hline
\end{tabular}
\caption{Haloes from the Aquarius simulation analysed in this work. 
  The left column shows the simulation name, encoding the halo (A to D),
  and the resolution level (2 and 4);  the corresponding particle masses 
  are shown in the right column, where $h = 0.73$.}
   \label{table:haloes}
\end{table} 

%%%%%%%%%%%%%%%%%%%%%%%%%%%%%%%%%%%%%%%%%%%%%%%%%%%%%%%%%%%%%%%%%%%%%%%%%%%%%%%%%%%%%%%%%%%
\subsection{Subhalo Finders} \label{finder}

Several subhalo finders have been applied on the Aquarius haloes.  
Each of them delivers a different subhalo catalogue, due to their specific
numerical techniques.
In order to minimize differences when comparing subhalo samples from different finders,
a common post-processing pipeline has been applied.
A detailed analysis of these subhalo catalogues can be found in \citet{Onions_2012}, 
whereas additional studies of the same data
have been published in several recent papers resulting from the 
"Subhalo Finder Comparison Project"  \citep[e.g.][]{Onions_2013, Knebe_2013, Elahi_2013, Pujol_2013}.

\revision{Eleven subhalo finders participated in the comparison project.
However, we present only results for the Amiga Halo Finder \citep[AHF;][]{Knollmann_2009}, 
 \subfind\ \citep{Springel_2001},  \rockstar\  \citep{rockstar_2013}, and the STructure Finder \citep[STF, also known as VELOCIraptor;][]{Elahi_2011}, since only
 those where run at high resolution on all four Aquarius haloes, analysed in this work, due to their computational efficiency.}
For further details of the different halo-finding algorithms,  we refer the reader to the corresponding code papers.

%%%%%%%%%%%%%%%%%%%%%%%%%%%%%%%%%%%%%%%%%%%%%%%%%%%%%%%%%%%%%%%%%%%%%%%%%%%%%%%%%%%%%%%%%%%
%%%%%%%%%%%%%%%%%%%%%%%%%%%%%%%%%%%%%%%%%%%%%%%%%%%%%%%%%%%%%%%%%%%%%%%%%%%%%%%%%%%%%%%%%%%
\section{Shape measurement}
\label{sec:shape_measurement}
To quantify the shape of dark matter haloes and their subhalo populations we aim to
approximate their isodensity contours with ellipsoids. This allows us to quantify
shapes in terms of the axis ratios $q=b/a$ and $s=c/a$, where $a$, $b$ and $c$
are the major, intermediate and minor axis of the ellipsoid respectively. In the following we will
refer to these axis ratios as shape parameters.
Several methods have been introduced in the literature to approximate isodensity contours with ellipsoids,
while most of them deliver similar results \citep[see][]{Allgood_2006, VeraCiro_2011}.
To evaluate the shape parameters we follow \citet{Dubinski_1991}, by
calculating the eigenvalues of the reduced moment of inertia
\begin{equation}
 I_{i,j}=\sum_n^N \frac{r_{i,n}r_{j,n}}{r_{1,n}^{2}+r_{2,n}^{2}+r_{3,n}^{2}},
\end{equation}
where  $\mathbf{r_n} = (r_{1,n}, r_{2,n}, r_{3,n})$ is the position of the $n^{th}$ object
with respect to the halo centre and N is the number of objects (either dark matter particles or subhaloes) that fulfill $|\mathbf{r_n}| \le 250$ $h^{-1}kpc$.
This radius was chosen to enclose all MW satellites, while it is larger than $r_{\rm 200}$ of the considered Aquarius haloes, given by \citet{Springel_2008}.
We chose the reduced instead of the usual moment of inertia to prevent objects with large distances to the halo centre
from dominating the measurements.

The subhalo masses are set to unity to be unaffected by uncertainties in their determination
and to simplify the comparison with observations.
Assuming  ellipsoidal distributions, the square root of the largest, intermediate and smallest eigenvalue corresponds
to the absolute value of the major, intermediate and minor axis respectively, that is, 
$(a, b, c) = \sqrt{(\lambda_1,\lambda_2,\lambda_3)}$, with $\lambda_3 \leq \lambda_2 \leq\lambda_1$.

Contrary to what is often done, we do not measure shapes iteratively.
In the iterative shape measurement particles that reside within the fitted ellipsoid
are used to remeasure the shape, while for the initial measurement all particles are taken into account. This
procedure is repeated until the shape parameters converge.
We do not employ this method because during the iteration objects are excluded from the analysis which
complicates the interpretation of shape measurements of, for instance, the N most massive subhaloes.
Furthermore we found that $q$ and $s$ sometimes converge to zero during the iteration, when only a few particles are analysed.
\citet{Bailin_2005} showed  that,
if shape measurement is performed without iteration,
selecting particles in a sphere biases the measured shape parameters towards spherical results.
We have verified these results using the \Aq{A }dark matter field.
However, in this study we are not interested in the exact values of the shape parameters, but more in how the
shapes of the subhalo populations are related to those of their hosts.

Besides the method for deriving the axis ratios, shape measurements can depend on the definition of the halo centre.
A possible offset between the centre of mass of the host halo and the one of the full matter distribution, caused by massive substructure,
can introduce an artificial triaxiality, especially when shapes are measured within the central part of a halo.
Moreover, an offset can indicate an unrelaxed condition of the halo according to results of \citet{Power_2012}, who
found that the distance between the most bound particle and the centre of mass is correlated with the virial ratio of the halo.
To find the centre of the host halo we apply the following procedure: i) particles within the distance $R/n$ around the centre of
mass are selected to recalculate a new centre of mass, where $n$ is initially set to unity and $R$ is the largest particle distance
to the centre of mass at the highest resolution level.  ii) Step i) is repeated $n = 100$
times, while n is increased by one in every iteration step.
Visual inspection confirmed that the centre determined by this procedure corresponds to the centre of
the host halo.
We use the host halo centre for the shape measurement of both - the dark matter field,
as well as the subhalo population. This approach is chosen because it is easily applicable to observational
data sets, assuming that the most luminous galaxy (e.g. the MW) resides in the centre of the host halo \citep[see e.g.][]{Vale_04}.
This latter assumption  might be wrong in some cases  \citep[see e.g.][]{Skibba_2011, Trujillo_11}.
%%%%%%%%%%%%%%%%%%%%%%%%%%%%%%%%%%%%%%%%%%%%%%%%%%%%%%%%%%%%%%%%%%%%%%%%%%%%%%%%%%%%%%%%%%%
%%%%%%%%%%%%%%%%%%%%%%%%%%%%%%%%%%%%%%%%%%%%%%%%%%%%%%%%%%%%%%%%%%%%%%%%%%%%%%%%%%%%%%%%%%%
\section{results}\label{sec:results}

%*************************************************************************************%
\begin{figure*}
\centering
{\includegraphics[width=8.6  cm, angle=270]{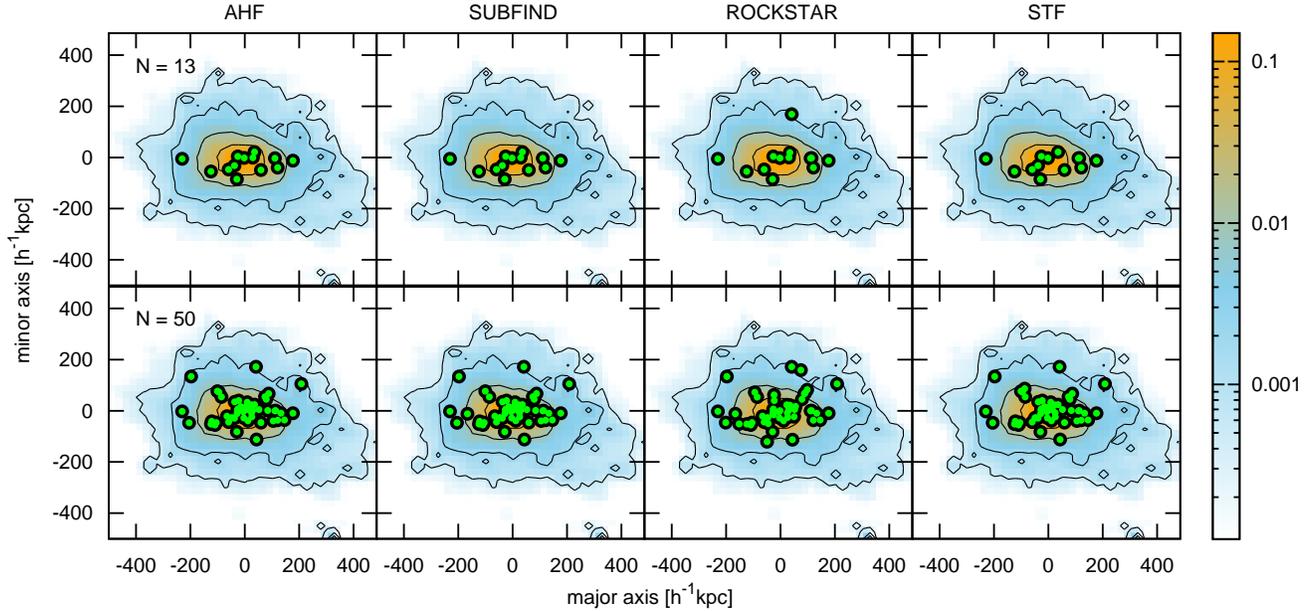}}
\caption
{Matter distribution of the \Aq{A} halo together with the population of the $13$ (upper panels) and the
$50$ (bottom panels) subhaloes with the highest {\vmax} values.  Colours shown on the right bar denote
the normalised density of dark matter  at the resolution level 4 in a $100$ $h^{-1}kpc$ slice through the
halo centre perpendicular to the  intermediate axis measured at $250$ $h^{-1}kpc$. The black-rimmed
green dots denote positions of the $N$  \Aq{A} subhaloes with highest {\vmax} values within a sphere of
$250$ $h^{-1}kpc$ identified by the subhalo finders {\ahf}, {\subfind}, {\rockstar} and {\stf} at resolution
level 2.
Note that in the case of {\rockstar} and {\stf} only $12$ of the $13$ highest
subhaloes are visible since one is obscured by other subhaloes.} \label{fig:results_1}
\end{figure*}
%*************************************************************************************%

In this section we present our shape measurements of the Aquarius subhalo populations and compare
them to the shape of the underlying dark matter field as well as to observational data.

In the whole analysis we quantify subhalo masses with the values of the current maximum rotational
velocity {\vmax} with which dark matter particles orbit around the subhalo centre. \citet{Kravtsov_2003}
found that these {\vmax} values are correlated with the subhalo masses, even during tidal stripping.
\citet{Onions_2012} demonstrated  that when {\vmax} is used as mass proxy the mass function of
subhaloes obtained from different halo finders are in better agreement with each other than for
$M_{200}$. This might be explained by {\vmax} measurements being less dependent on uncertain
definitions, such as the subhalo boundaries.

%%%%%%%%%%%%%%%%%%%%%%%%%%%%%%%%%%%%%%%%%%%%%%%%%%%%%%%%%%%%%%%%%%%%%%%%%%%%%%%%%%%%%%%%%%%
\subsection{Distribution of subhalo populations identified by different finders}\label{sec:results_1}

\revision{To gain a visual impression of how subhaloes are distributed around the centre of
their host halo we display the \Aq{A} dark matter field together with its subhalo population in
Figure \ref{fig:results_1}.}
The halo is rotated so that the intermediate axis of the ellipsoid approximating the dark matter isodensity contours stands
perpendicular to the plane of the figure. This ellipsoid is fitted to the  dark matter distribution within a $250$ $h^{-1}kpc$
sphere around the halo centre.
The coloured area denotes the normalised projected density of the dark matter particles at the second
lowest level of resolution (level $4$) within a $100$ $h^{-1}kpc$ slice through the halo centre and parallel
to the intermediate axis. The full dark matter field at different resolution levels was not available for our
analysis, whereas we do not expect this limitation to affect our conclusions as discussed
in Section \ref{sec:results_2}.
The black-rimmed green dots denote the positions of the $13$ and $50$ highest {\vmax} subhaloes
\footnote{The selection of the $13$ highest {\vmax} subhaloes is motivated by the set of $13$
MW satellites that are brighter than $M_V=-8.5$, listed in Table \ref{table:MWsats}. This set of
satellites consists mostly of objects  detected before the SDSS analysis (the "classical" satellites).
The number of highest {\vmax} subhaloes in the larger sample corresponds to the $50$ MW
satellites, predicted by \citet{Simon_2007} for a completely observed sky.}
(top and bottom panels, respectively), identified by the subhalo finders {\ahf}, {\subfind}, {\rockstar}
and {\stf}  at the second highest level of resolution (level $2$).

We find that the subhalo distributions identified by the different halo finders differ only by a few
objects.  Especially the results of {\ahf} and {\subfind} agree very well with each other.
We also find that the $13$ \Aq{A} subhaloes with the highest {\vmax}
values are roughly aligned with the major axis of the dark matter field, while their distribution
is flatter. An alignment in the same direction is also apparent for the $50$ subhaloes with
the highest {\vmax} values, while the shape of their distribution is more isotropic and,
therefore, in better agreement with the shape of the underlying dark matter field.

\revision{A physical explanation for the flattening of the massive subhalo population might
be that massive objects maintain information about their infall direction for a longer time
due to their large moment of inertia, while the orbits of low mass subhaloes randomise
faster after infall. If subhaloes are accreted anisotropically, the triaxiality of a given subhalo
distribution might therefore increase the more massive objects are contained in the samples.
Furthermore tidal forces can strip of material from subhaloes, while they orbit around the
centre of their host. This might introduce an additional correlation between the subhalo
mass and its orbit orientation with respect to the infall direction. Another reason might be
that dynamical friction causes massive subhaloes to sink faster to the host center.
For this reason they might more prone to be totaly disrupted before developing
circular orbits, as suggested by \citet{Gill_2004}.
However, due to the small
number of massive subhaloes the flatness of their distribution might be an effect of random
sampling. We attempt to disentangle such a systematic effect from anisotropic accretion
as reasons for the flatness in Section \ref{sec:results_2}.}

%%%%%%%%%%%%%%%%%%%%%%%%%%%%%%%%%%%%%%%%%%%%%%%%%%%%%%%%%%%%%%%%%%%%%%%%%%%%%%%%%%%%%%%%%%%
\subsection{Flatness of the subhalo population}\label{sec:results_2}

\revision{Focusing the analysis on the flattening of the massive subhalo population with respect to the dark matter field
we measure their shapes using results from the four haloes \Aq{(A-D)}.
To quantify shapes we follow the procedure described in Section \ref{sec:shape_measurement} by 
determining the largest, intermediate and smallest eigenvalues of the reduced
moment of inertia (a, b, c respectively). The latter is constructed from objects
(either dark matter particles or subhaloes) that reside within a $250$ $h^{-1}kpc$ sphere
around the halo centre. The shape can then be quantified by the shape parameters 
$q=b/a$ and $s=c/a$ which correspond to the axis ratios of ellipsoids approximating
the dark matter and subhalo distributions. Note that according to this definition,
the extreme cases $q=s=1$, $s \ll q = 1$ and $q = a \ll 1$ correspond,  respectively,
to a sphere, a disc and a filament.}

%%%%%%%%%%%%%%%%%%%%%%%%%%%%%%%%%%%%%%%%%%%%%%%%%%%%%%%%%%%%%%%%%%%%%%%%%%%%%%%%%%%%%%%%%%%
\subsubsection{Systematic bias in shape measurements from small numbers of tracers}
\label{sec:results_2_1}

%*************************************************************************************%
\begin{figure*}
\centering\includegraphics[width=8.75 cm, angle=270]{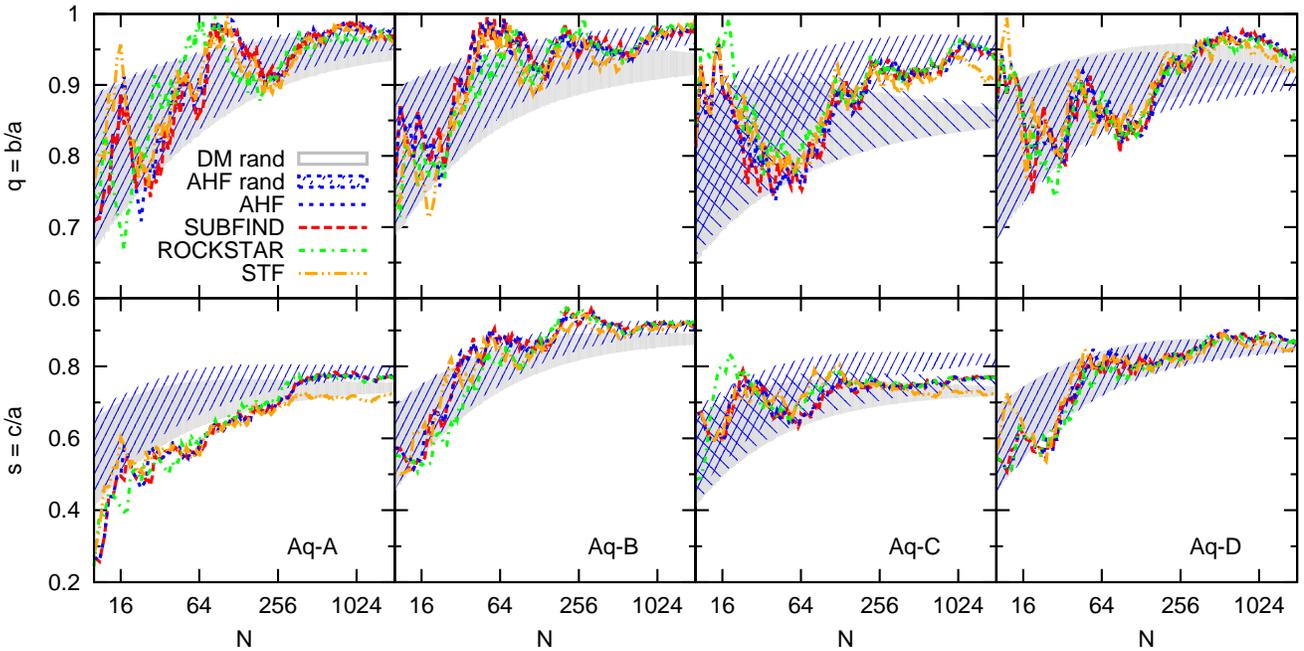}
\caption
{Shape parameters $q$ (upper panels) and $s$ (bottom panels) of \Aq{(A-D)} subhalo populations (panels from left to right) at the resolution level $2$ for the $N$  subhaloes with the highest {\vmax} values within $250$ $h^{-1}kpc$
around the halo centre. The blue hatched (grey) areas result from random selections of $N$ subhaloes (dark matter particles at the resolution level $4$) regardless of their mass.
They denote $68\%$ confidence levels derived from $1000$ realisations of such random selections. 
As explained in Subsection \ref{sec:results_2_1}, the wide blue hatched area in the panel for the \Aq{C} halo results from
a validation test to explain the difference between shapes from dark matter and subhalo random samples.}
\vspace{-5 pt}
\label{fig:results_4}
\end{figure*}
%*************************************************************************************%

A systematic bias in the shape measurement towards triaxiality from low numbers of shape tracing points (in our case subhaloes)
is expected since any discreteness noise will produce a nonzero ellipticity even from a perfectly spherical object.
In the extreme cases of two and three tracer points any density field will be classified as a filament
and disc respectively. This known effect \citep[e.g.][]{Paz_2006, vanDaalen_2012, Joachimi_2013} might be the reason
why we measure smaller values of the shape parameters $q$ and  $s$ for the distribution of the most massive
subhaloes than for larger sets including less massive subhaloes or the dark matter field.

We show in Fig.~\ref{fig:results_4} the values of $q$ and $s$ measured from the distributions of the $N$  most massive 
\Aq{(A-D)} subhaloes using {\vmax} as a proxy for mass. Different coloured line types stand for the results of the subhalo
samples obtained from the four halo finders that were run at the resolution level 2 on all haloes.
\revision{The shape parameters show a  $\Delta q \simeq \Delta s \simeq 0.1$ scatter when roughly more than $100$ subhaloes are
used for the measurement. For smaller numbers of subhaloes results from different finders show a stronger scatter of around
$\Delta q \simeq \Delta s \simeq 0.2$. A more detailed analysis based on results from ten subhalo finders, run at all five resolution
levels of the \Aq{A} halo, revealed a stronger scatter for a larger variety of finders, which tends to increase at low resolution.
At the lowest resolution level $5$ we find a scatter of roughly $\Delta q \simeq \Delta s \simeq 0.4$ for samples consisting of the $13$ highest {\vmax} subhaloes,
partly because of the larger number of finders run at low resolution.
This finding suggests that the shapes of subhalo populations consisting of the highest {\vmax} objects in MW-like host haloes, measured in currently high
resolution cosmological simulations, strongly depend on the subhalo finder employed to identify substructures.}

As indicated by the visual impression from Fig. \ref{fig:results_1} the shape parameters increase (shapes become more spherical) if more subhaloes
with lower mass (lower {\vmax} values) are used for the shape measurement. This effect is apparent for all four haloes  \Aq{(A-D)}, while the
measurements also show strong fluctuations for different numbers of most massive subhaloes.
We mentioned Section \ref{sec:results_1} that the decrease of shape parameters for the $N$ most massive subhaloes might be caused by physical effects,
such as anisotropic infall, but also by a systematic effect resulting from small numbers of subhaloes in these samples.

To distinguish between the two possible reasons for the flatness, we generate random samples of $N$ {\ahf} subhaloes that are randomly selected
from all detected subhaloes within a given halo, independently of their {\vmax} value.
If the shape parameters of the distribution of the $N$ most massive subhaloes are significantly smaller than the values obtained from many random samples,
then this would indicate a physical reason for the flatness.
Systematics from small numbers of subhaloes should not cause differences between shapes from the two types of samples, since both consist of the
of the same number of objects.
We generate $1000$ random samples, each with $N$ {\ahf} subhaloes, from which we obtain median values of the shape parameters and $68\%$
probability regions as indicator for the scatter. The latter is shown as a blue hatched region in Fig.~\ref{fig:results_4}.
We find that the shape parameters  $q$ and $s$ of the distributions of the $N$ most massive subhaloes roughly lie within the central
$68 \%$ region of $q$ and $s$ measurements derived from random samples. We have tested that randomly sampling subhaloes
identified by different subhalo finders leads to the same result.
In most cases, the increased triaxiality (smaller shape parameters) of the $N$ most massive subhaloes with respect to the dark
matter field can therefore be explained by a systematic bias in the shape measurement due to small numbers of objects. An exception
of this result is the $s$ parameter of \Aq{A}, which is significantly smaller for samples of the $N$ most massive subhaloes than the expected
value from random sampling. We will discuss this finding in Section \ref{sec:results_2_2} in more detail.

\revision{
The results in Fig.~\ref{fig:results_4} also show that the variation between shape measurements for subhalo populations at level $2$ from different finders
is smaller than the noise in the measurement, expected from random sampling. The aforementioned scatter at low resolution is comparable
to the noise from random sampling.}

\revision{When the number of considered shape tracers is above $1000$ the shape measurements start to converge to certain values of $q$ and $s$.}
Studying the shapes of artificial haloes we find that the convergence of the shape parameters for large numbers of tracer particles depends
on the halo shape. The number of particles necessary to reach a certain precision in the shape measurement increases the closer a
halo is to spherical symmetry, while in that case roughly $1000$ tracer particles are needed for a $10\%$ precision (see Fig.~\ref{fig:app_1}
in Appendix~\ref{sec:app1}). Note that this result might depend on the exact method used for the shape measurement. Although similar results
are reported by authors using different definitions of the moment of inertia \citep[e.g.][]{Joachimi_2013}, further study might be worthwhile to
explore how heavily different shape measurements  are affected by discreteness noise.

Comparing the shapes of the distributions of the $N$ most massive subhaloes with the dark matter halo shape, taking into account the bias in the
measurements from low numbers of objects, we also generate $1000$ samples of $N$ randomly selected dark matter particles from each Aquarius halo at the resolution level $4$, 
instead of using subhaloes.
\revision{As mentioned before, only the resolution level $4$ was available for this analysis. \citet{VeraCiro_2011} demonstrated that
their \Aq{A} dark matter halo shape parameters at the five resolution levels differ by less than $4\%$ from each other,
when the measurement is performed at radii as large as those used in our study ($250$ $h^{-1}kpc$).
Their shape parameters measured from dark matter particles within $250$ $h^{-1}kpc$ around the halo centre are smaller than ours,
in part because we measure shapes without iteration. We are able to roughly reproduce their results if we measure the dark matter shapes
iteratively as well. Moreover, the difference between our measurement of the shape parameter $s$ derived with and without iteration is
consistent with results reported by \citet{Bailin_2005}.}

\revision{Besides measuring dark matter shapes without iteration a further difference to the measurements of \citet{VeraCiro_2011} is that we do not
exclude particles that are members of subhaloes from the analysis.}
We do not follow this approach, as the relation between the shape of the subhalo population
and the full dark matter shape is closer to observational information, as dark matter halo shapes might be traced for example by X-ray or lensing signals, which result from the full
mass distribution. The $68\%$ probability regions of the $1000$  shape measurements are shown as grey regions in the same figure.

If the random samples of subhaloes and dark matter particles consist only of a few objects, both types of random samples show similar distributions of the shape parameters. 
However, for larger numbers of objects, the shape parameters of subhalo and dark matter random samples do not converge to the same values. We can explain
this result if we bear in mind that  each substructure is represented by just one single object (the subhalo) in case of the subhalo random samples, but as several
objects (the dark matter particles of the subhalo) in case of the dark matter random samples. This feature might introduce an artificial anisotropy in the dark matter
moment of inertia which is reflected by smaller shape parameters. In this sense, the fact that the difference between shape measurements from random subhalo
and dark matter particle samples is stronger for the \Aq{C} halo might be related to the fact that this halo has more high-mass substructures, as it can be seen
in Fig.~3 from \citet{Springel_2008}.

To gain a better understanding of this effect, we perform a simple test that consists in giving more weight to the massive subhaloes 
during the random selection. For this purpose, we simply write subhaloes with large values of $v_{\rm max}$ ($v_{\rm max}> 100\,km/s$) $2000$ times in the subhalo
catalogue and repeat the shape measurement from random samples. Following this simple approach, we can roughly reproduce the results 
from the dark matter random sampling, as shown by the additional wide blue hatched area included in the panel for the \Aq{C} halo in Fig.~\ref{fig:results_4}. 
For the sake of clarity, we only show the result of this test for the \Aq{C} halo, although we have checked that we obtain similar results for the rest of Aquarius haloes.

This same effect could have been tested by excluding dark matter particles in subhaloes from the measurement.

%%%%%%%%%%%%%%%%%%%%%%%%%%%%%%%%%%%%%%%%%%%%%%%%%%%%%%%%%%%%%%%%%%%%%%%%%%%%%%%%%%%%%%%%%%%
\subsubsection{Anisotropic infall}\label{sec:results_2_2}

We found in the previous section that the value of $s$ measured from the $N$ most massive \Aq{A} subhaloes is significantly
below the expected value from randomly sampling subhaloes (Fig.~\ref{fig:results_4}). It is therefore unlikely that the flatness
of the \Aq{A}  population of massive subhaloes can be explained by random sampling. Since this effect is independent of the employed subhalo finder
we show, for the sake of clarity, here and in the following analysis only results derived from {\ahf} subhalo populations. We have tested that our conclusions do not change when employing different finders.

\begin{figure}
\centering
\begin{subfigure}
\centering\includegraphics[width=7.0 cm, angle=270]{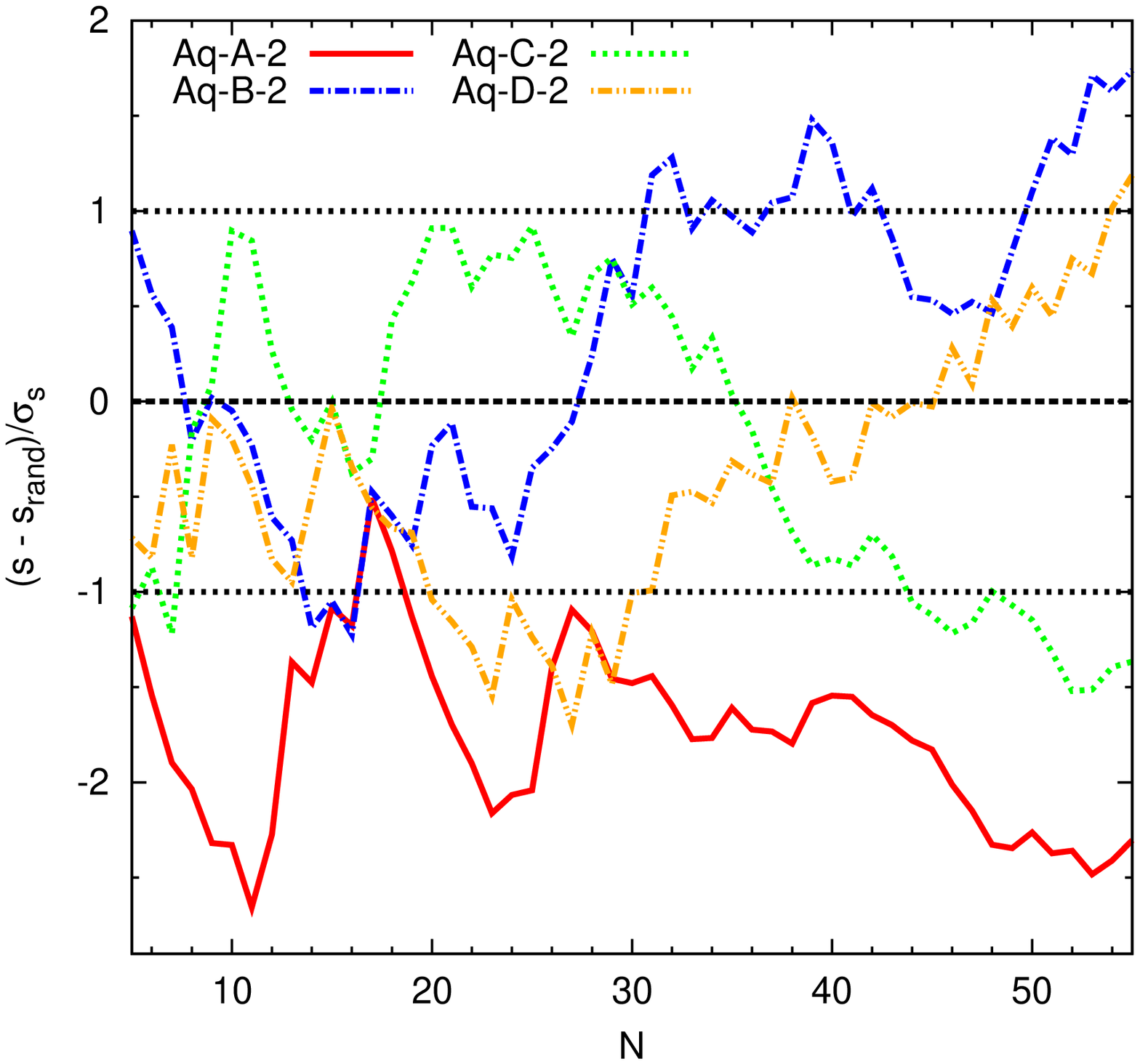}
\label{fig:results_5a}
\end{subfigure}
\begin{subfigure}
\centering\includegraphics[width=7.0 cm, angle=270]{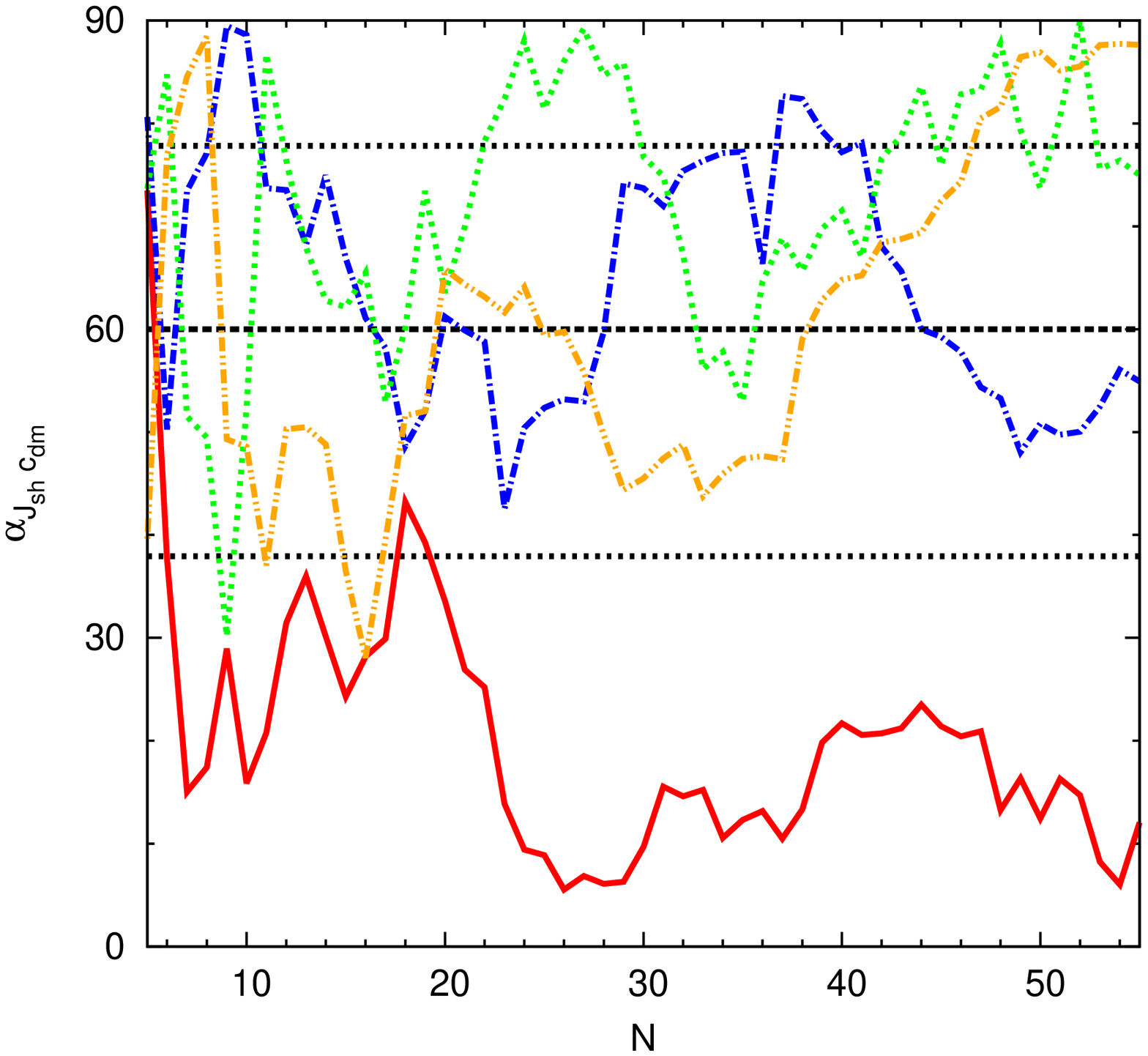}
\label{fig:results_5b}
\end{subfigure}
	\caption{\emph{Top panel:} Difference between the shape parameter $s$ calculated from the $N$ most highest {\vmax} {\ahf} subhaloes and the
	median value of $1000$ shape measurements from $N$  randomly selected {\ahf} subhaloes ($s_{\text{rand}}$) divided by the $1\sigma$
	errors, shown as a hatched area in Fig.~\ref{fig:results_4}. \emph{Bottom panel:} Angle between the total angular momentum of the 
	subhalo population (with subhalo masses set to unity) and the minor axis of the dark matter field. The dashed and dotted lines
	indicate the mean angles $\pm 1\sigma$ deviations from an isotropic distribution respectively.}
\label{fig:results_5}
\end{figure}

Focusing the analysis on the flatness of the massive \Aq{A} subhaloes we show in the top panel of  Fig.~\ref{fig:results_5} the significance of the deviation between the shape parameter $s$ of the $N$ most massive
\Aq{(A-D)} subhaloes and the median values obtained from randomly sampling subhaloes. We define this significance as the ratio of the deviation and the error obtained from randomly sampling subhaloes,
shown as blue hatched areas in Fig. \ref{fig:results_4}.  Again, we see that the \Aq{A} subhalo population is significantly flatter than expected from random sampling, while the flatness of massive subhalo
populations in the other haloes lies within the expected range.

In order to understand this particular result, we study, in the following, anisotropic infall as an additional source of flattening.
A proper way  to study anisotropic infall would be to follow the subhalo positions through different time steps of the simulation, like for example \citet{Libeskind_2005}.
However,  since this data is not available within the present comparison project, we investigate the orbit orientation of subhaloes given by the direction of their angular
momenta with respect to the dark matter field of the corresponding host haloes.
\revision{This approach is motivated by the following consideration: if the $N$ highest
{\vmax} subhaloes do not accidentally lie in a plane, then also their orbits should lie within the same plane, possibly because they entered the host halo from a similar direction. 
The angular momentum $\mathbf{J}$ of this subhalo population should therefore be aligned with the minor axis of the ellipsoid fitted to the dark matter halo,
if the orbital plane is contained within the plane defined by the major and intermediate axes of the dark matter ellipsoid (hereafter referred to as the major plane).}

We calculate the total angular momentum of subhalo populations as
\begin{equation}
\mathbf{J} = \sum_n^N \mathbf{J_n}\, ,
\end{equation}
where $\mathbf{J_n} = \mathbf{r_n}\times \mathbf{v_n}$ is the angular momentum of the $n^{th}$ subhalo,
$\mathbf{r_n}$ and $\mathbf{v_n}$ are the corresponding position and velocity vectors
with respect to the halo centre and its velocity, and $N$ stands for  the number of highest {\vmax} subhaloes.
As in the computation of the moment of inertia, all  subhalo masses are set to unity. 

In the bottom panel of Fig.~\ref{fig:results_5} we show the angle between the
minor axis of the dark matter field of each host halo {\Aq{(A-D)}} and the total angular momentum of the corresponding subhalo populations 
when we take samples consisting of the $N$ highest {\vmax} subhaloes.
We find that this angle is smaller for \Aq{A} than for the other haloes, indicating
a stronger alignment  between the subhalo populations angular momentum and the dark matter minor axis.
If we associate this stronger alignment with the effects of anisotropic infall,
this result suggests that anisotropic infall can slightly increase the flattening of the subhalo populations, 
while the main reason for the flattening is the bias in the shape measurement due to small numbers of objects.

Note that the individual subhaloes do not contribute equally to the total angular momentum, since they are weighted by their distance
and velocity with respect to the halo centre. Consequently, few objects with high velocities, large distances to the
centre, or both can dominate the measurement. We therefore study in the following the flattening of subhalo populations
and their rotational support with quantities that are defined for individual objects.

%*************************************************************************************%
%*************************************************************************************%

%*************************************************************************************%
\begin{figure}
\centering
\begin{subfigure}
\centering\includegraphics[width=7.0 cm, angle=270]{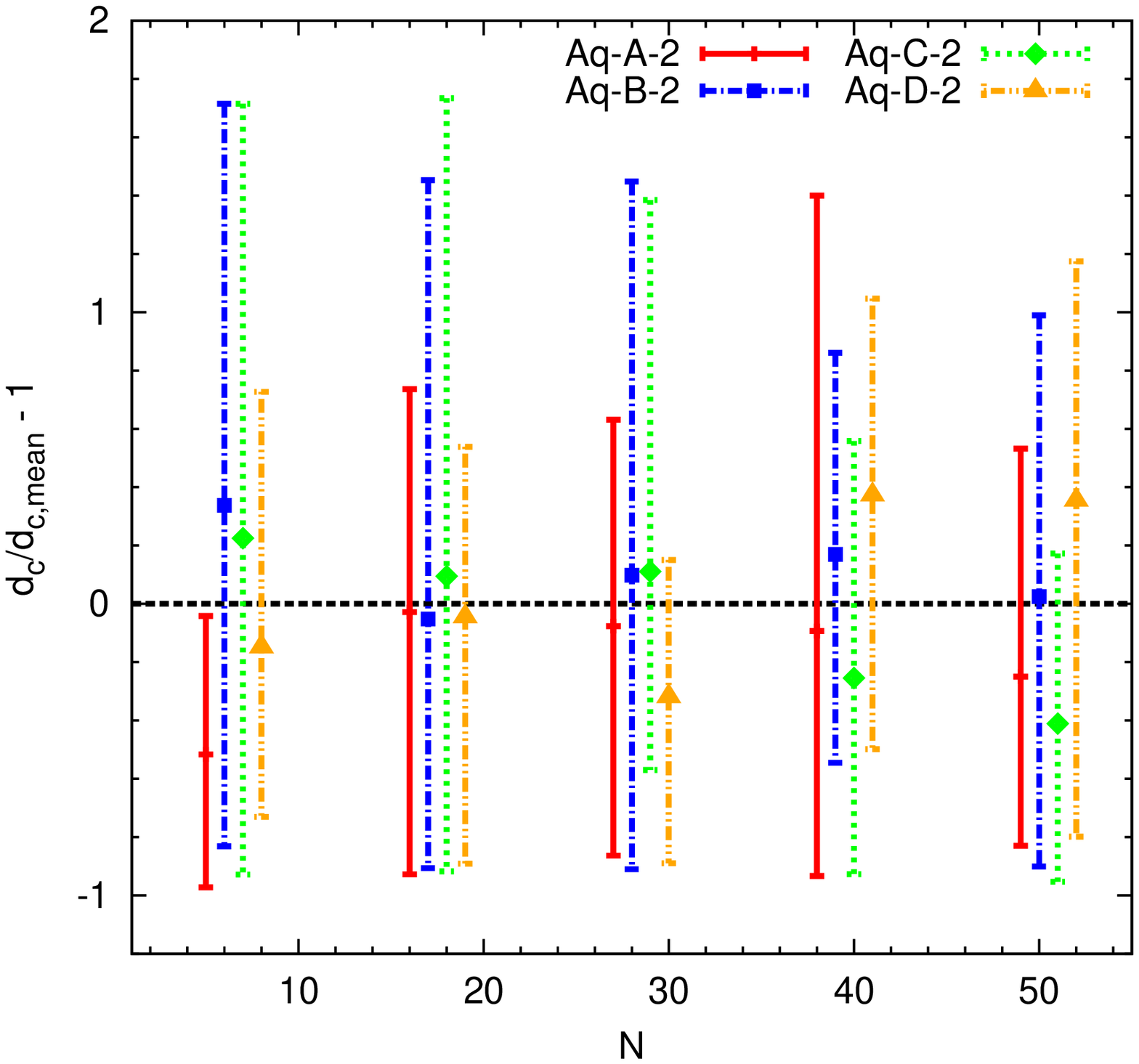}
\label{fig:results_6a}
\end{subfigure}
\begin{subfigure}
\centering\includegraphics[width=7.0 cm, angle=270]{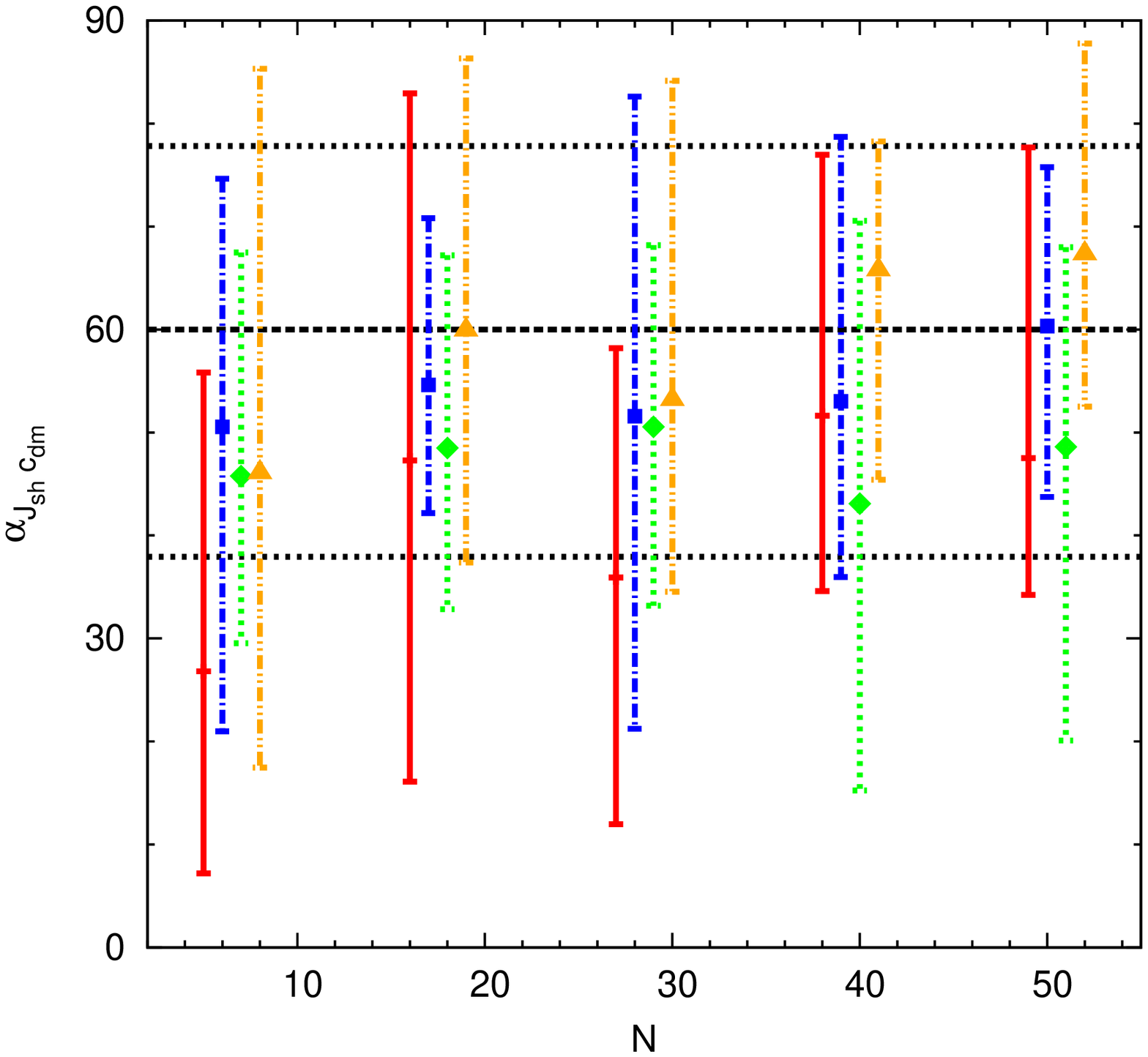}
\label{fig:results_6b}
\end{subfigure}
	\caption{\emph{Top panel:} Distance, $d_c$, of each subhalo to the plane spanned by the major and the intermediate axis of the dark matter
	moment of inertia in units relative to the mean distance to the same plane computed  over the whole subhalo population. Symbols show the mean
	$d_c/d_{c,\text{mean}}-1$ over $\text{N}\pm5$ {\ahf} subhaloes with the highest {\vmax} values. Error bars enclose the central $68\%$ of the distribution in each bin.
	\emph{Bottom panel:} Angle between angular momentum of each subhalo and the minor axis of the dark matter moment of inertia. Binning and 
	error bars are calculated in the same way as in the top panel. The dashed and dotted lines indicate the mean angles $\pm 1\sigma$ deviations
	from an isotropic distribution respectively.}
\label{fig:results_6}
\end{figure}
%*************************************************************************************%

We start with a complementary method to study the degree of flattening of the subhalo population which consists in analysing the distribution
of distances of the individual subhaloes to the major plane of the dark matter ellipsoid.
From the subhalo distances to the dark matter major plane, $d_{c}$ we derive the mean over the
whole subhalo population, $d_{c,\text{mean}}$. We show the mean relative distance $d_{c}/d_{c,\text{mean}}$ as computed in
bins of $11$ subhaloes with the highest {\vmax} values in the top panel of Fig.~\ref{fig:results_6}. For each bin
we also show error bars that enclose the $68\%$ of the values around the mean of the distribution. 
In general, we find that the subhaloes of \Aq{(B-D)} with the highest
{\vmax} values do not lie significantly closer to the major plane of the dark matter field than the mean over the whole population.
The situation is slightly different for \Aq{A}, whose most massive subhaloes seem to be closer to the major plane, indicating therefore a flatter distribution.
This result is consistent with those from shape measurements shown in Fig.~\ref{fig:results_5}, where we derived
cumulative instead of differential values of subhalo populations with the highest {\vmax} values.

In order to verify rotational support, we measure for each subhalo separately the angle between the angular momentum 
and the minor axis of the ellipsoid fitted to the dark matter field as $\alpha_n = \arccos(\hat{J}_n \cdot \hat{c}_{dm})$. 
Using the same bins of  $11$ subhaloes sorted by {\vmax} as in the top panel of Fig.~\ref{fig:results_6}, 
we now show, in the bottom panel of the same figure, the mean values of $\alpha_n$ in each bin together with the  
the error bars which enclose the central $68\%$ of the distribution.
In agreement with our previous results, we find that the \Aq{A} subhalo population, which
is significantly flattened, is also stronger aligned with the minor axis of the dark matter ellipsoid than it is the case for the other haloes.
These findings point towards an scenario in which the flatness of the \Aq{A} subhalo population can be partially explained
by the effects of  anisotropic infall.

Results of \citet{VeraCiro_2011} also support this scenario (see their Fig. 8). These authors show that material falling into the \Aq{A} halo has
a strong dipole component between  $11 - 8$ $Gyr$ lookback time,
indicating filamentary accretion. The dipole components that they find for the other haloes are smaller, as material
was accreted more isotropically. This result also confirms the reports by \citet{Lovell11} who also found indications for anisotropic accretion
by studying the alignment between the angular momenta of Aquarius subhaloes (as detected by \subfind) and their dark matter host halo.
More specifically,  \citet{Lovell11}  found that the distribution of angular momentum directions of subhaloes with 
the $11$ largest progenitors had the strongest dipolarity in the case of \Aq{A}.

%%%%%%%%%%%%%%%%%%%%%%%%%%%%%%%%%%%%%%%%%%%%%%%%%%%%%%%%%%%%%%%%%%%%%%%%%%%%%%%%%%%%%%%%%%%
\subsection{Comparing shapes of simulated subhalo populations  with those of MW satellites}\label{sec:results_2_3}

In this Section we compare the shapes of the Aquarius subhalo populations to those of the
MW satellite population consisting of up to $27$ objects within $250$ $h^{-1}kpc$ around the galactic centre.
As we have ranked subhaloes by their {\vmax} values, we now use V-band luminosities to rank the satellite galaxies.
Our comparison with MW data is therefore based on the assumption that the satellite galaxies with the brightest V-band magnitudes ($M_V$) live in the subhaloes
with highest {\vmax} values. As we use {\vmax} as a proxy for the total subhalo mass,  $M_V$ can be used to quantify the stellar mass of a satellite galaxy \citep{Baraffe_1998}.
A weak correlation between the stellar mass of satellite galaxies in MW like halos and their {\vmax} values can be found in simulations \citep{Brooks_2012, Zolotov_2012,
Rodriguez_2013, Reddick_2013}.
On the contrary, \citet{Libeskind_2005} found that the stellar masses of semi-analytical model satellites are not correlated with the corresponding total subhalo
mass at $z=0$, but with the total mass of the main progenitor, as suggested by \citep{Conroy_2006}. However, their employed model does not include tidal
stripping of the stellar mass component.
Despite the existing uncertainties, we assume that {\vmax} is a convenient property for comparing results of the most massive subhaloes to those derived from
the observed most luminous MW satellites.
Note that all masses are set to unity in the definition of the moment of inertia from which we measure the shapes.
The positions and V-band luminosities of the $27$ MW satellites, taken from \citet{McConnachie_2012}, are shown
in the Appendix \ref{sec:app3}.

\begin{figure*}
\centering\includegraphics[width=11.5 cm, angle=270]{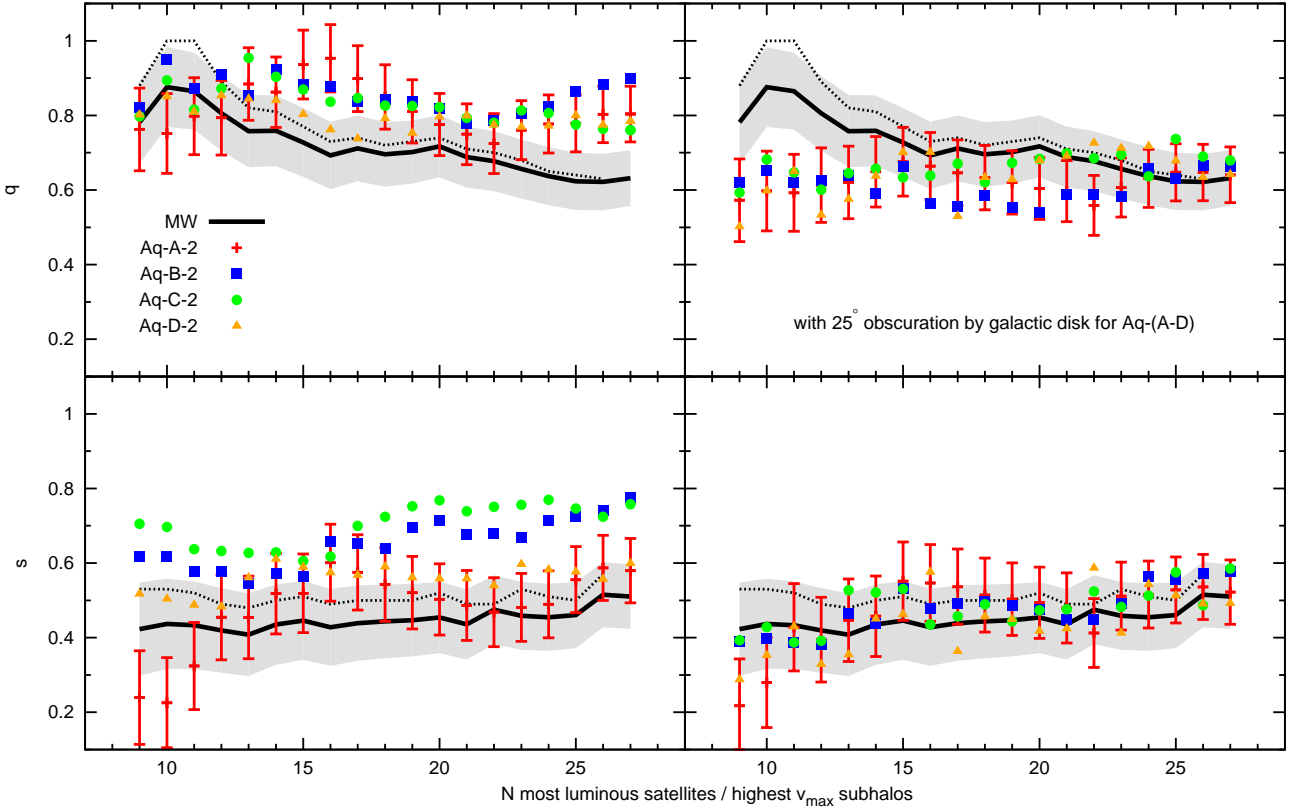}
\caption{Shape parameters of the MW satellite distribution consisting in the $N$  objects with highest V-band luminosities (black lines).
These shapes are compared to those of Aquarius subhalo populations consisting in the $N$  objects with highest {\vmax} values
identified by the {\ahf} subhalo finder. The left panel shows results for the highest {\vmax} subhaloes in the full sky. The right panel shows results for selecting
subhaloes in a light cone with $130$ degree opening angle, assuming $25$ degree obscuration by the galactic disc.
The coloured symbols denote the median values from $1000$ realisations with randomly oriented light cones.
The grey area and the error bars show the standard deviations estimated as described in the Appendix \ref{sec:app2}.
Black dashed lines show shape parameters from the MW data measured with a method that decreases systematic bias from low numbers of tracers,
as described in Appendix \ref{sec:app1}.}
\label{fig:results_7}
\end{figure*}

The shape parameters measured from satellite populations consisting of the $N$  brightest objects are shown as black lines in the left panel of Fig.~\ref{fig:results_7}.
We also show the shape measurements of the \Aq{(A-D)} subhalo populations, consisting of the $N$  objects with highest {\vmax} values.
We find that the absolute values of the shape parameters $q$ measured from the satellite populations are similar to the results from simulations for samples that consist of the
roughly twelve or less most massive objects. When fainter satellites are included to the measurement we find a decrease of $q$ while the Aquarius results
remain roughly constant, resulting in lower values for $q$ in observations than in simulations. 
The increase of the $q$  from \Aq{(A-D)} with the number of objects, which we have seen for large numbers in Fig.~\ref{fig:results_4}, is not apparent
when the analysis is restricted to the $27$ highest {\vmax} subhaloes.
In the case of the $s$ parameter we find that the observational values tend to be below the results from simulations except for samples of the highest mass {\Aq{A}}
subhaloes. Our values of $s$ for the population of the twelve most luminous satellites is in rough agreement with those reported by \citet{Starkenbur_2013}.
The $s$ measurements for the twelve highest {\vmax} subhaloes roughly agree with results that these authors obtain from randomly sampling
satellites with $M_V<-8.5$ from semi-analytic models of galaxy formation imposed on the Aquarius simulation.

The stronger discrepancies between the shape parameters from observed satellite and simulated subhalo populations for larger samples
might be explained by an angular selection effect, which has a stronger impact on faint objects.
We expect selection effects to be twofold. On the one hand, galactic obscuration prevents us from detecting objects that lie close to the galactic plane. 
On the other hand, the sky is observed with an inhomogeneous intensity, due to the limited area of surveys. Both types of selection effects should affect stronger faint, as opposed to bright, objects.
To study the effect of angular selection on the shape measurement we follow the procedure described by \citet{Wang_2013} by measuring the shapes of the highest {\vmax} {\Aq{(A-D)}}
subhalo population residing within a light cone with an opening angle of $\alpha=180-2\alpha_{\text{obs}}$, where $\alpha_{\text{obs}}$ is the angle of obscuration with respect to the galactic plane.
Assuming that the orientation of this synthetic galactic plane is arbitrary, we conduct $1000$ shape measurements with randomly oriented galactic planes.
Orienting the galactic plane randomly is motivated by reports on the weak alignment of observed blue late-type galaxies and disc galaxies from hydrodynamic simulations
with their surrounding matter distribution at large scales \citep{zhang_2013, Joachimi_2013, sales_2012, hahn_2010}.
The mean results of these $1000$ shape measurements are shown in the right panel of Fig.~\ref{fig:results_7}. We find that both $q$ and $s$ are
decreased when galactic obscuration is taken into account. Using an obscuration angle of $\alpha_{\text{obs}}=25^\circ$ we find that, for the population
of all $27$ satellites, the observational results for the shape parameters agree with those from simulations.
If the analysis is restricted to the $12$ brightest satellites, the observed shape parameters are higher than those
from simulations. This indicates that faint samples are stronger effected by obscuration, as we have initially guessed.
However, angular selection effects are probably more complicated than in our model. The obscuration angle of $\alpha_{\text{obs}} =25^\circ$ is larger than commonly
reported for the MW. In Fig.~\ref{fig:app_7} we see that the majority of satellites resides outside of this zone. This might be explained by a smaller obscuration
angle together with the angular limitation of the SDSS (both shown in \citealt{metz_2009}), which result in a larger effective obscuration angle.

The black dashed lines in Fig.~\ref{fig:results_7} show the shape parameters of satellite populations measured with a method that decreases the
bias in the shape measurement from low numbers of shape tracing objects. This method is described in Appendix \ref{sec:app1}.

%*************************************************************************************%
%*************************************************************************************%

%%%%%%%%%%%%%%%%%%%%%%%%%%%%%%%%%%%%%%%%%%%%%%%%%%%%%%%%%%%%%%%%%%%%%%%%%%%%%%%%%%%%%%%%%%%
%%%%%%%%%%%%%%%%%%%%%%%%%%%%%%%%%%%%%%%%%%%%%%%%%%%%%%%%%%%%%%%%%%%%%%%%%%%%%%%%%%%%%%%%%%%
\section{Summary and Conclusions}\label{sec:summary}

We have analysed four high resolution simulations of dark matter haloes with MW-like masses from the Aquarius project, \Aq{(A-D)}, +
to measure the shapes of their subhalo populations and compared them to those of the underlying dark matter distributions.
Throughout the analysis we used the {\vmax} values of the subhaloes as a proxy for the subhalo mass.

\revision{We find that the shapes of subhalo populations, consisting of the objects with the highest {\vmax} values show only a weak dependence
on the employed subhalo finder when the subhaloes are analysed at high mass resolution. This result changes for small samples at lower mass resolutions since shape
measurements for smaller subhalo samples are highly sensitive to changes of a few members. We therefore expect similar measurements, performed in currently high
resolution cosmological simulations (e.g. Millennium-II), strongly depend on the subhalo finder used to identify substructures.}

Our shapes measured from subhalo populations consisting of the $N$ subhaloes with highest {\vmax} values show that the subhalo distributions
are more triaxial when only objects with highest {\vmax} values are considered. This trend is apparent for all four halo finders run at the resolution level $2$ and all four Aquarius haloes.
Interpreting these results in terms of anisotropic accretion is delicate due to the fact that triaxiallity is likely to increase for samples with a few subhaloes.
As a $null$ test we conducted the same shape measurement for $N$ randomly selected subhaloes and dark matter particles. In all
cases we found a decrease of the shape parameters for small numbers of objects.
In three of the four analysed haloes the decrease of the shape parameters for the subhaloes with highest {\vmax} values is consistent with results from
random samples. For \Aq{A} the shape parameter $s$ is up to $40\%$ smaller than the median value from random sampling. This difference roughly
corresponds to a  $2\sigma$ deviation and indicates a flattening of the population of  high {\vmax} subhaloes due to anisotropic infall.

To further verify if the flattening of the high {\vmax} subhalo populations is not caused by random sampling, we studied the alignment of the angular momenta of
the subhalo populations with the ellipsoid of the dark matter field. We found that the highest {\vmax} \Aq{A} subhaloes are stronger aligned with the minor axis of the
dark matter ellipsoid than the \Aq{(B-D)} subhalo populations. This finding indicates that the orbits of the high {\vmax} \Aq{A} subhaloes tend to reside in the plane
spanned by the major and intermediate axis of the dark matter moment of inertia, which is in agreement with subhalo accretion along the host halo major axis.
Such an interpretation is consistent with the findings of \citet{VeraCiro_2011}, who showed that the matter accreted by the \Aq{A} halo
has a dipole component - a signal of filamentary infall. Also \citet{Lovell11}  report indications for anisotropic accretion by studying the alignment between
the angular momenta of Aquarius subhaloes and the dark matter field of their host halo.

After having analysed possible reasons for the increased triaxiallity of the highest {\vmax} Aquarius subhaloes, we compare the shape of their distributions with those of observed MW satellites.
In this regard, we found that the shape parameters derived for the distribution of the twelve satellites with the brightest V-band magnitudes lie within the range
expected from the four Aquarius haloes when using the same number of subhaloes with the highest {\vmax} values.
However, for larger samples including fainter satellites and lower {\vmax} subhaloes the shape parameters are below the results from simulations, especially the major to intermediate axis
ratio $q$. This discrepancy can be decreased by assuming an overly strong galactic obscuration of $25^\circ$, while in that case the shape parameters
for the high {\vmax} samples lie below the results from bright satellites samples. This demonstrates that the geometry of our observational window,
defined by the zone obscured by the galactic disc, and the geometry of the surveys mapping the sky can effect the shape measurements of the
satellite distribution, while we expect a stronger impact on samples including faint satellites.

Within the $\Lambda$CDM context, the strong triaxiality of the MW satellite distribution with respect to our results from simulations
can be interpreted as an result of anisotropic accretion, which would agree with the findings of \citet{libeskind11}.
\revision{A further reason for the strong flatness of the observed satellite distribution might be that the assumption that the most
luminous satellites live in the subhaloes with the highest mass at redshift zero  (using the maximum rotational velocity as mass proxy)
might be inadequate to mimic the results of galaxy formation \citep{Boylan-Kolchin_2011}. Simulations and observations suggest that the subhalo mass after infall and their stellar mass
are not or just weakly correlated, while we assume a monotonic relation for our comparison with MW data  \citep{Libeskind_2005, Brooks_2012, Rodriguez_2013}.
It might be necessary to simulate the evolution of dark matter together with the baryonic physics to produce thinner satellite discs \citep{Danowich_12}.}

Our analysis of systematic effects on the shape measurement revealed that hundreds of tracers are needed for obtaining reliable results. We conclude that even
if subhaloes, as potential hosts of satellite galaxies, follow the dark matter field of their host, they are to few for a reliable shape measurement of the MW
dark matter halo. The expected increase of the number of known satellites in future surveys \citep{Simon_2007} will probably not be sufficient to solve this problem.
However, we also found that shapes of subhalo populations in three of four MW-like host haloes are consistent with the shapes of the dark matter field,
if bias from the small number of tracers in the shape measurement is taken into account. The assumption that subhaloes trace the dark matter field of their host is therefore
in reasonable agreement with our results.

\revision{A similar analysis performed on larger host haloes with more massive subhaloes at different redshifts would allow for an improved comparison
with larger observational data sets from cosmological surveys.}

%------------------------- acknowledgements -------------------
\section*{Acknowledgements}
This paper was initiated at the ÓSubhaloes going NottsÓ workshop in Dovedale, UK, which was funded by the European Commissions
Framework Programme 7, through the Marie Curie Initial Training Network Cosmo- Comp (PITN-GA-2009-238356).
We wish to thank the Virgo Consortium for allowing the use of the Aquarius dataset and Adrian Jenkins for assisting with the data.

KH is supported by beca FI from Generalitat de Catalunya. He acknowledges Noam Libeskind for a fruitful discussion.
SP and HL acknowledge a fellowship from the European Commission's Framework
Programme 7, through the Marie Curie Initial Training Network CosmoComp (PITN-GA-2009-238356).
SP also acknowledges support  by the PRIN-INAF09 project ``Towards an Italian Network for Computational Cosmology'' and  by
{\it Spanish Ministerio de Ciencia e Innovaci\'on} (MICINN) (grants AYA2010-21322-C03-02 and CONSOLIDER2007-00050).
EG and KH acknowledge the {\it Spanish Ministerio de Ciencia e Innovaci\'on} (MICINN) project AYA2009-13936, Consolider-Ingenio CSD2007- 00060,
AK is supported by the {\it Ministerio de Econom\'ia y Competitividad} (MINECO) in Spain through grant AYA2012-31101 as well as the Consolider-Ingenio 2010 Programme of the {\it Spanish Ministerio de Ciencia e Innovaci\'on} (MICINN) under grant MultiDark CSD2009-00064. He also acknowledges support from the {\it Australian Research Council} (ARC) grants DP130100117 and DP140100198. He further thanks Felt for penelope tree.
SIM acknowledges the support of the STFC Studentship Enhancement Program (STEP).
YA receives financial support from project AYA2010-21887-C04-03 from the former {\it Ministerio de Ciencia e Innovaci\'on} (MICINN, Spain), as well as the Ram\'{o}n y Cajal programme (RyC-2011-09461), now managed by the {\it Ministerio de Econom\'ia y Competitividad}  (fiercely cutting back on the Spanish scientific infrastructure).
PSB is supported by a Giacconi Fellowship through the Space Telescope Science Institute, which is supported through NASA contract NAS5-26555.
MN thanks the Sir John Templeton Foundation for support through a New Frontiers and Astronomy and Cosmology grant.

The authors contributed in the following ways to this
paper: KH and SP undertook this project.
KH performed the analysis presented and wrote the paper together with SP.
KH is a PhD student supervised by EG. FRP, AK,
EG, HL, JO and SIM contributed with useful discussions and 
with the organisation of  the workshop at which this
study was initiated. They designed the comparison study
and planned and organised the data. The other authors provided
data and had the opportunity to proof, read and
comment on the paper.

%---------------------------- references  ---------------------------
\bibliographystyle{mnbst}
\bibliography{hoffmannetal_rev1_arxiv.bib}

%---------------------------- appendix ---------------------------
\appendix 

%%%%%%%%%%%%%%%%%%%%%%%%%%%%%%%%%%%%%%%%%%%%%%%%%%%%%%%%%%%%%%%%%%%%%%%%%%%%%%%%%%%%%%%%%%%
%%%%%%%%%%%%%%%%%%%%%%%%%%%%%%%%%%%%%%%%%%%%%%%%%%%%%%%%%%%%%%%%%%%%%%%%%%%%%%%%%%%%%%%%%%%
\section{Improved shape measurement}\label{sec:app1}

\begin{figure}
\centering\includegraphics[width=8.5 cm, angle=0]{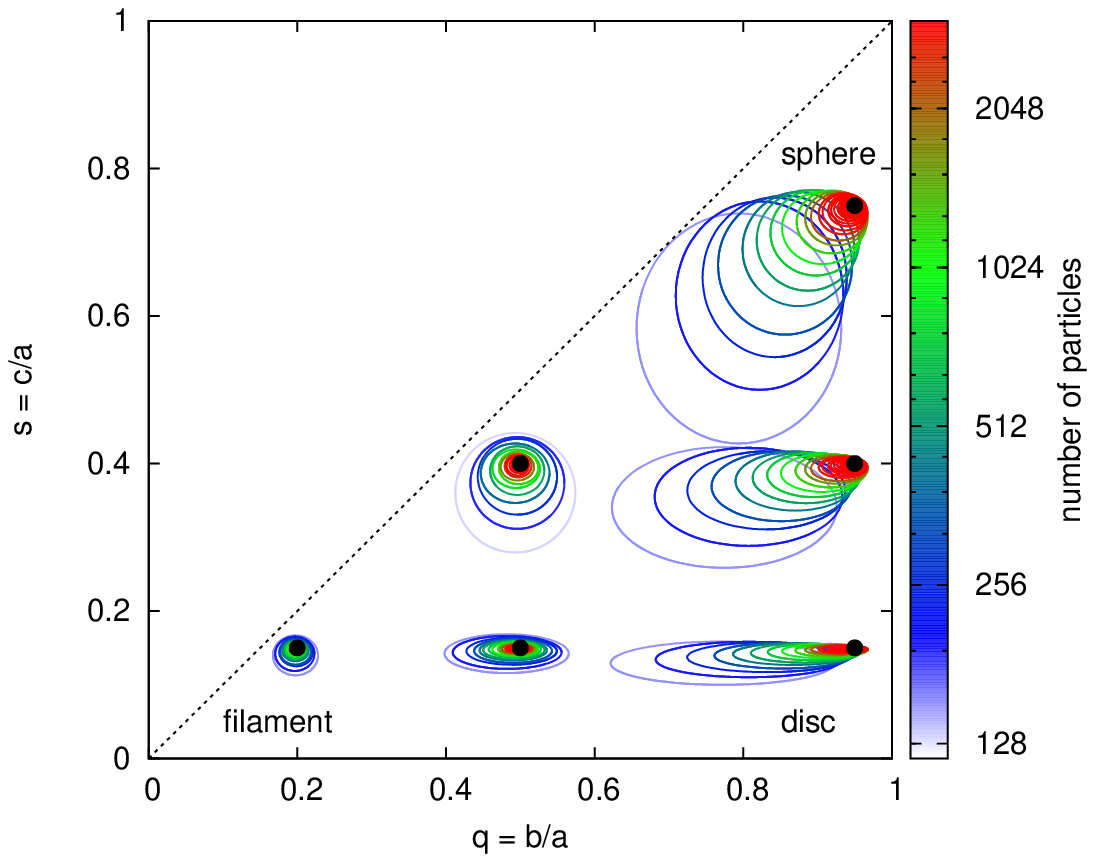}
\caption{Measured shape parameters of artificial particle distributions with given input shapes, marked by black dots. The centre of the ellipses are the mean measurements of $q$ and $s$ derived from $10000$ random realisations of the particle distributions. The major and minor axes denote the standard deviation of $q$
and $s$ measurements, respectively. The colour of the ellipses denotes the number of random particles used in each of the $10000$ realisations.}
\label{fig:app_1}
\end{figure}

To study in detail the bias induced in the shape measurement due to low numbers of tracer points, we perform the following test.
We build mock dark matter haloes by generating particle distributions characterised by a given ellipticity and density profile.
Using $10000$ random realisations of such artificial haloes formed by $N$tracers (representing subhaloes or dark matter particles), 
we measure the mean shape parameters $q$ and $s$
and the corresponding standard deviations for different sets of input shapes as a function of $N$. The results of this analysis, shown in Fig.~\ref{fig:app_1}, reveal that
around $1000$ tracer particles are required to ensure that the measured shape parameters deviate by less than $10\%$ from the true (input) values.
The bias in the shape measurement due to low tracer points is stronger if the shape parameters of the artificial haloes are close to unity, while results are almost unbiased if the shape parameters are below $0.2$.

%***************************************************************%
%***************************************************************%

\begin{figure}
\centering
\includegraphics[width=7 cm, angle=270]{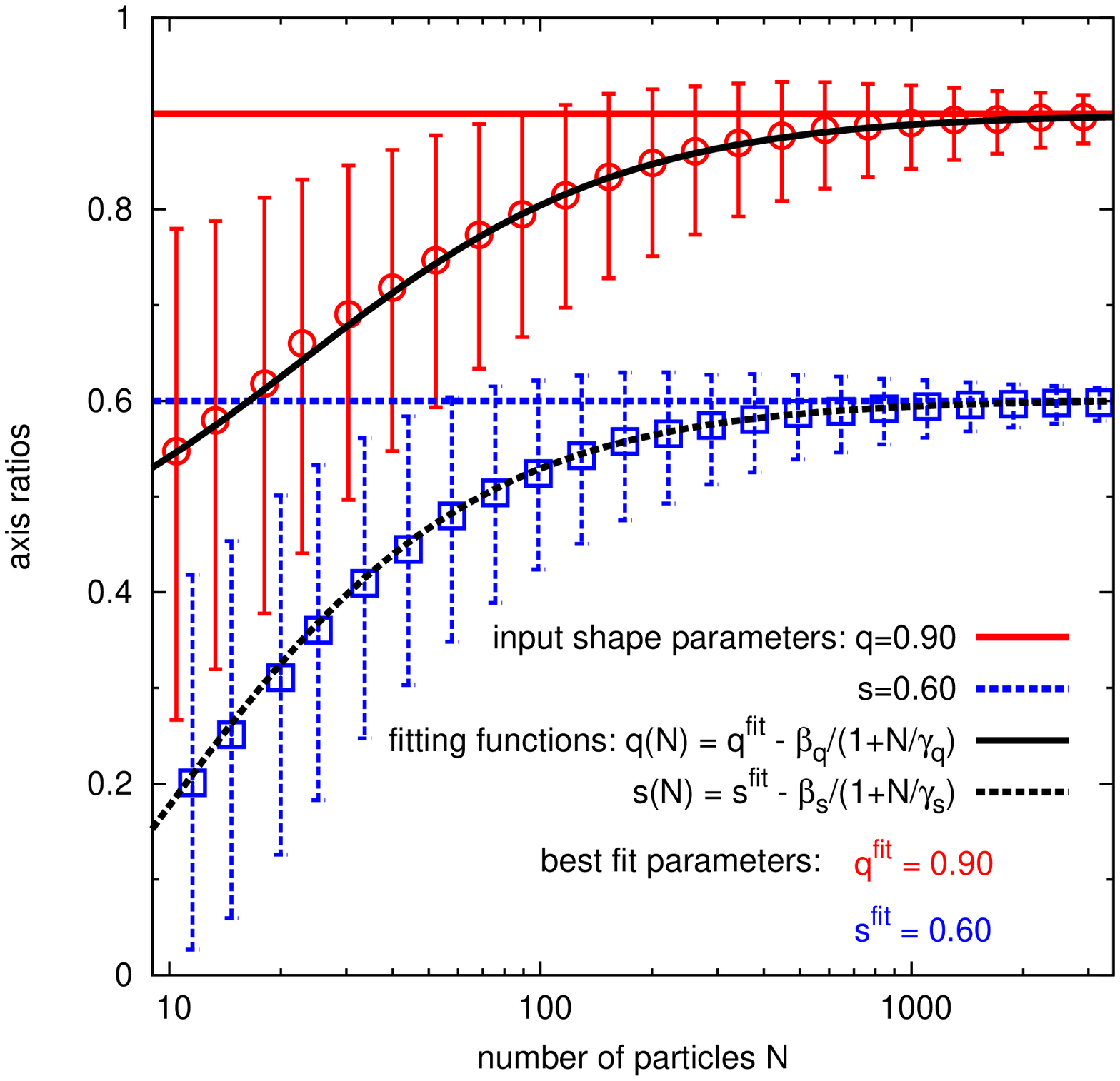}
\caption{Shape parameters of artificial particle distributions (mimicking a halo with $q=0.9$ and $s=0.6$) 
as a function of the number of tracer particles $N$.
The shape parameters, used to generate the particle distributions, are shown as the coloured horizontal lines. The symbols
show the median and central $68\%$ values obtained from $10000$ random realisations. Solid and dashed black lines show
fits of Eq.~\ref{eq:shapefit} to the measurements of $q$ and $s$, respectively.}
\label{fig:app_2}
\end{figure}

In Fig.~\ref{fig:app_2} we show the measured shape parameters of a mock halo 
with input $q=0.9$ and $s= 0.6$ as a function of the number of tracer particles, $N$.
By trial and error we find that the shape measurements $q(N)$ and $s(N)$ can be described 
with a function of the form
\begin{equation}
q(N )=q_{\text{fit}} - \frac{\beta}{1+N /\gamma},
\label{eq:shapefit}
\end{equation}
where $q_{\text{fit}}$, $\beta$ and $\gamma$ are free parameters. The expression for $s(N)$ is analogous to the one for $q(N)$. 
For large number counts $q(N)$ and $s(N)$ converge, respectively, to $q_{\text{fit}}$ and $s_{\text{fit}}$, which correspond to the 
true shape parameters of the halo.

%***************************************************************%
%***************************************************************%

\begin{figure}
\centering\includegraphics[width=14 cm, angle=270]{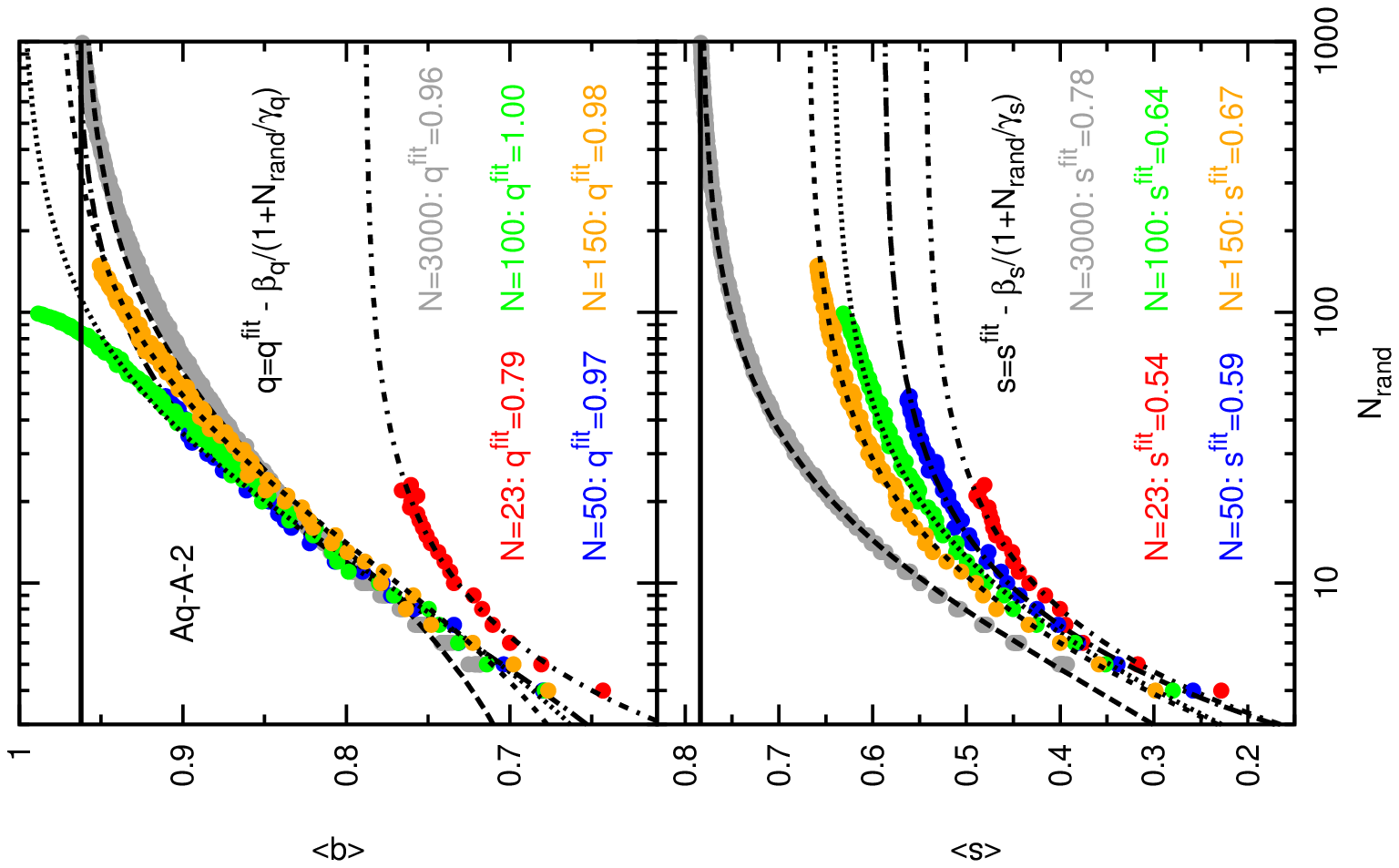}
\caption{Mean shape parameters derived from $1000$ random samples consisting of $N_{ \text{rand}}$ subhaloes (coloured symbols).
The samples are drawn from the $N${\Aq{A}} subhaloes with the highest {\vmax} values without replacement. The black dashed lines show
the fitting function given by Eq.~\ref{eq:shapefit} with the best fit values for $q_{\text{fit}}$ and $s_{\text{fit}}$ displayed in the figure.
The best fit values derived from the  $N=3000$  highest {\vmax} subhaloes are shown as black solid lines.}
\label{fig:app_3}
\end{figure}

These fitting functions open the possibility to decrease the bias in the shape measurement from low number 
counts using the following procedure.
If we consider the case of the subhalo populations, we randomly sample groups of subhaloes, as much as 
possible, without replacement. These samples are drawn from the $N$highest {\vmax} subhaloes, since
in real life only the most massive (or luminous) satellite galaxies are observed. The mean shape parameters 
of such groups are then measured as a function of the number of group members, $N_{\text{rand}}$. 
By fitting Eq.~\ref{eq:shapefit} to the measured $q(N_{\text{rand}})$ and $s(N_{ \text{rand}})$, we can derive
the parameters $q_{\text{fit}}$ and $s_{\text{fit}}$ as shown in Fig.~\ref{fig:app_3}.

%***************************************************************%
%***************************************************************%

\begin{figure}
\centering
\includegraphics[width=13.0 cm, angle=270]{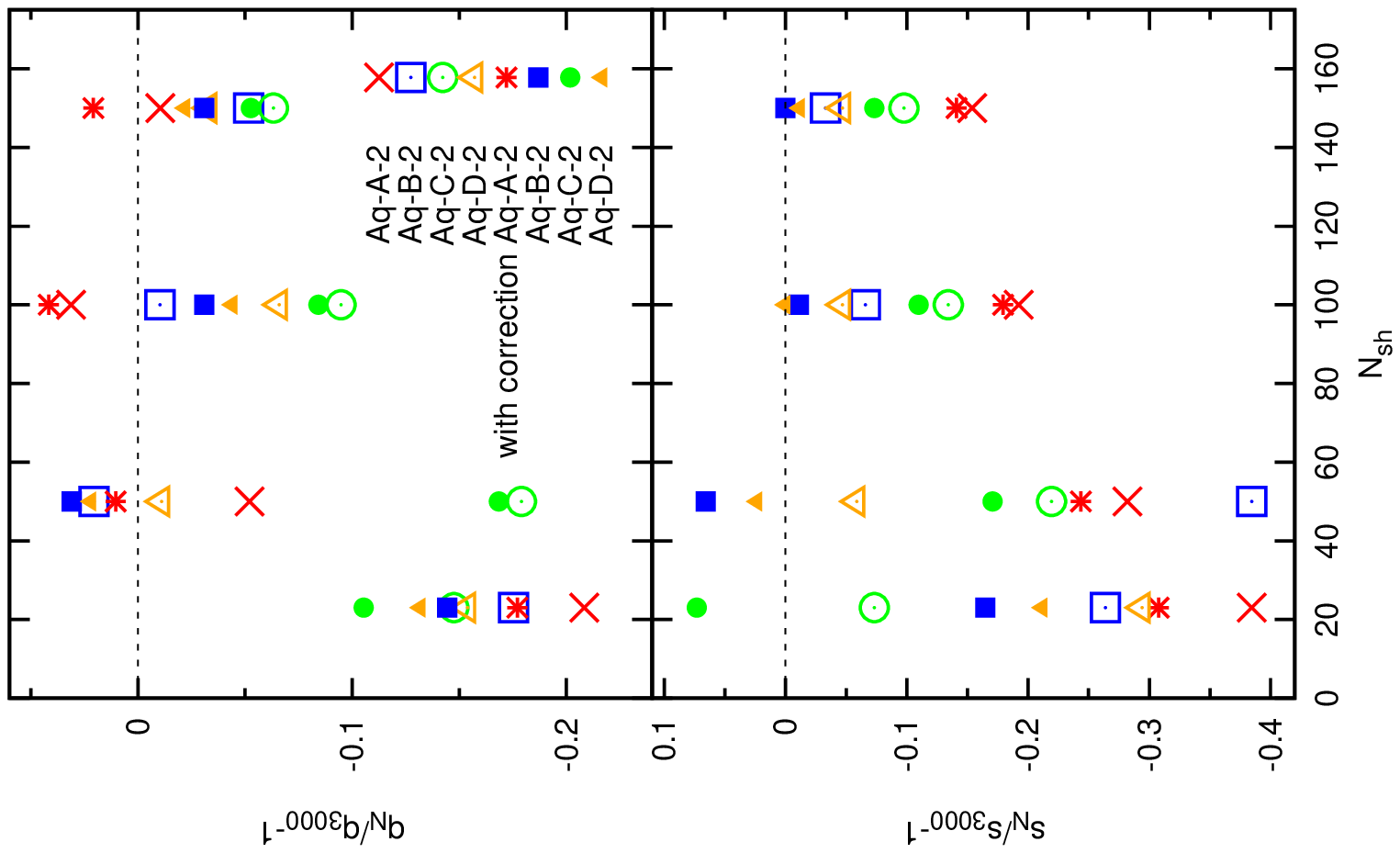}
\caption{Relative difference between the shape parameters $q$ and $s$ measured from the $N_{\text{sh}} $ and $3000$ \Aq{(A-D)}
subhaloes with the highest {\vmax} values. Open symbols show direct $q$ and $s$ measurements, whereas closed symbols
represent measurements corrected as described in the text and demonstrated in Fig.~\ref{fig:app_3}.}
\label{fig:app_4}
\end{figure}

These fitting parameters tend to lie closer to the shape parameters derived from 
the $3000$ subhaloes with the highest {\vmax} values than the shape parameters
measured using the $N<3000$ most massive subhaloes. 
Indeed, as it is shown in Fig.~\ref{fig:app_4}, for the $23$ highest {\vmax} subhaloes in \Aq{(A-D)}, the difference between
$q(N)$ and $q(3000)$ decreases by about $5 \%$, whereas in the case of $s(N)$ the deviation from $s(3000)$ is about 
$10 \%$ smaller if this correction procedure is applied.

%***************************************************************%
%***************************************************************%

%***************************************************************%
\begin{figure}
\centering
\includegraphics[width=12.5 cm, angle=270]{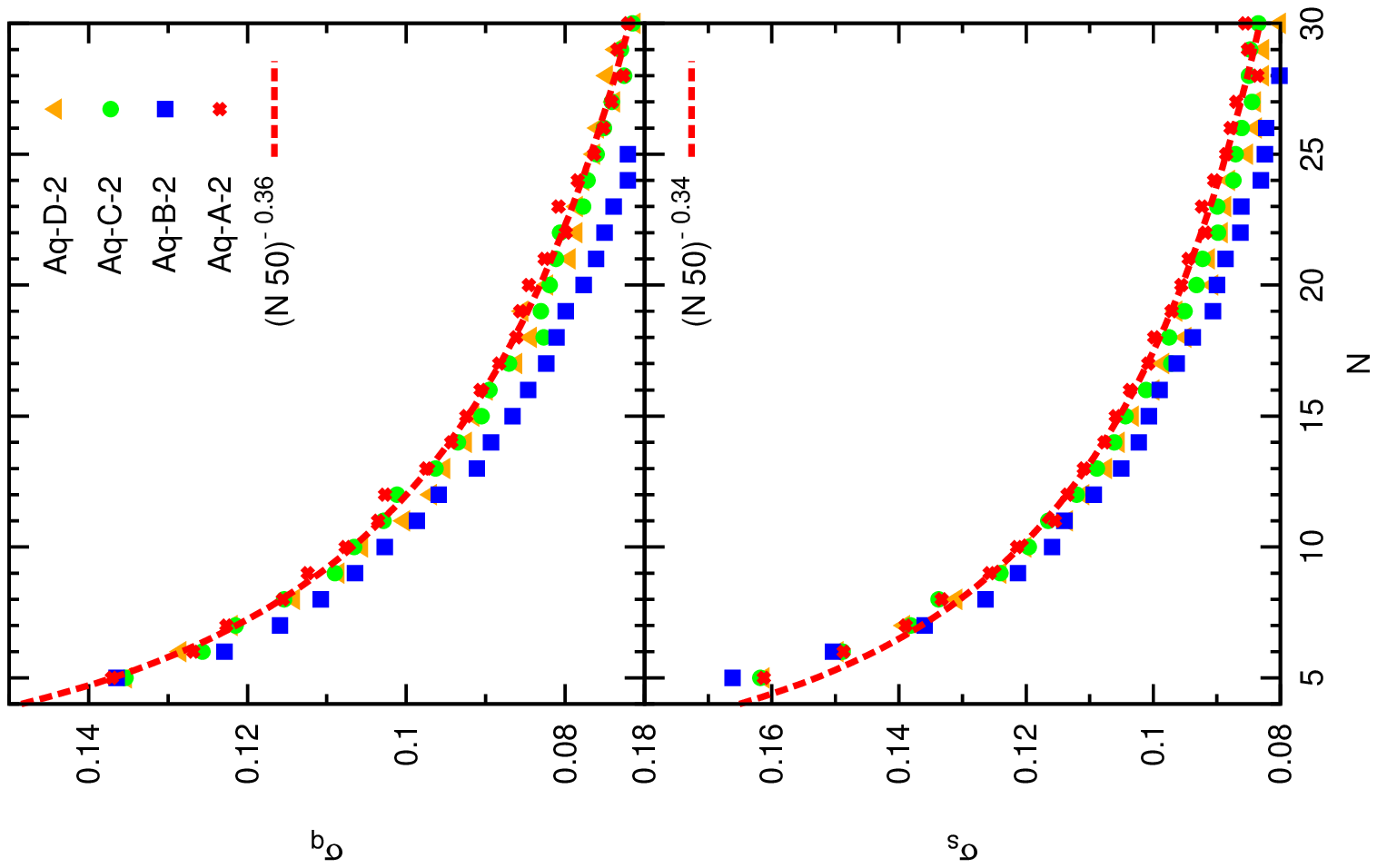}
\caption{Symbols denote the standard deviation of the shape measurement from randomly selected \ahf\  
subhaloes, shown as a blue shaded area in Fig.~\ref{fig:results_4}.
The red dashed lines are a power law approximation of the \Aq{A} results, used as an error estimation in Fig.~\ref{fig:results_6}.}
\label{fig:app_6}
\end{figure}
%***************************************************************%

%%%%%%%%%%%%%%%%%%%%%%%%%%%%%%%%%%%%%%%%%%%%%%%%%%%%%%%%%%%%%%%%%%%%%%%%%%%%%%%%%%%%%%%%%%%
%%%%%%%%%%%%%%%%%%%%%%%%%%%%%%%%%%%%%%%%%%%%%%%%%%%%%%%%%%%%%%%%%%%%%%%%%%%%%%%%%%%%%%%%%%%
\section{Error estimations for the shape measurement}\label{sec:app2}

We estimate the error of the shape measurements from the observational data and highest {\vmax} {\Aq{A}} subhaloes
shown in Fig.~\ref{fig:results_7} using the results from random sampling Aquarius  subhaloes independently of their {\vmax}
values. The standard deviation derived from sampling $N$\ahf\ subhaloes $10000$ times are shown in Fig.~\ref{fig:app_6}.
We find that these results can be approximated by a power law shown in the same figure. The error estimate applied to the
observational data is the same power law that describes the standard deviation from the {\Aq{A}} results.

%%%%%%%%%%%%%%%%%%%%%%%%%%%%%%%%%%%%%%%%%%%%%%%%%%%%%%%%%%%%%%%%%%%%%%%%%%%%%%%%%%%%%%%%%%%
%%%%%%%%%%%%%%%%%%%%%%%%%%%%%%%%%%%%%%%%%%%%%%%%%%%%%%%%%%%%%%%%%%%%%%%%%%%%%%%%%%%%%%%%%%%
\section{Milky Way satellites}\label{sec:app3}

\begin{figure*}
\centering\includegraphics[width=6.3 cm, angle=270]{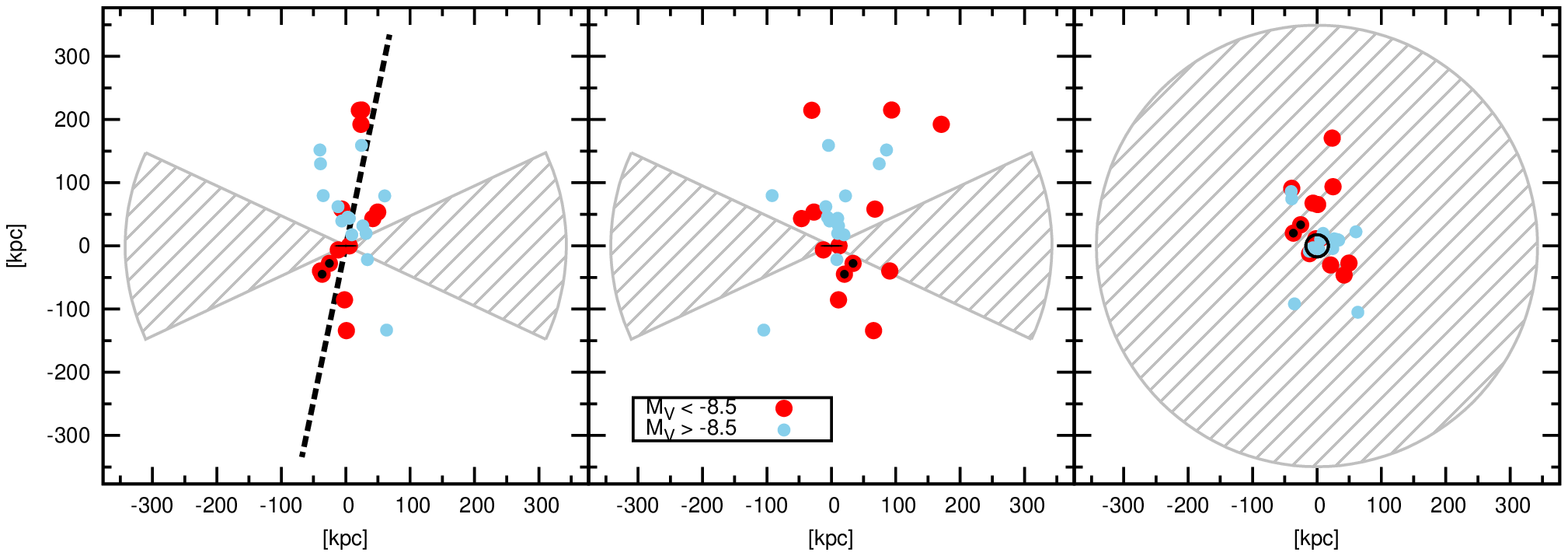}
\caption{Distribution of satellites from Table \ref{table:MWsats} around the MW centre.
Left and central panel show edge on views on the galactic disc.
Coordinates are rotated, so that the best fit plane, spanned by the major  and intermediate axis of the ellipsoid fitted to all $27$ objects, is edge on in the left panel. 
The central panel shows the same distribution while the perspective is rotated by $90^{\circ}$ around the galactic pole. The right panel shows a face on view
on the galactic disc. Red and blue dots denote, respectively, satellites with V-band magnitudes brighter and fainter than $M_{V}=-8.5$. The shaded area is the assumed galactic
obscuration zone used to calculate the shapes for Aquarius subhalo samples shown in the right panel of Fig.~\ref{fig:results_7}. Its radial size of $250$ $h^{-1}kpc$,
with $h=0.73$ corresponds to the size of the spherical window within which we select Aquarius subhaloes for the analysis. The obscuration angle of $25^\circ$
is about twice as large as the value usually assumed. \citet{metz_2009} show a similar figure with a smaller galactic obscuration zone together with the projected
SDSS light cone, which encloses all faint satellites. The extension of the galactic disc is roughly indicated by black bars in the left and central panel and an open circle
in the right panel. Small and large Magellanic Clouds are marked by black dots for orientation. }
\label{fig:app_7}
\end{figure*}

For comparing the shapes of the Aquarius subhalo distributions to MW data we use the positions and absolute V-band magnitudes of $27$
satellite galaxies from \citet{McConnachie_2012}, presented in Table~\ref{table:MWsats}. The distribution of these satellites 
around the centre of the MW is shown in Fig.~\ref{fig:app_7}.

The cartesian coordinates of each satellite are calculated from the heliocentric galactic angles $l$, $b$ and $r_{\text{GC}}$ as follows:
\begin{align*}
&x = r_{\text{GC}} \cos(b) \cos(l) - d_{\text{GC}} \\
&y = r_{\text{GC}} \cos(b) \sin(l) \\
&z = r_{\text{GC}} \sin(b) \, ,
\end{align*}
where the conversion to cartesian coordinates is based on a distance to the galactic centre of $d_{\text{GC}} = 8.4$ $kpc$ \citep{Ghez_2008}.
Our cartesian coordinates slightly differ from those given in \citet{Pawlowski_2013}
due to differences in the angles and in the $d_{\text{GC}}$ used for the coordinate conversion. Note that the observational coordinates
are given $kpc$ to be consistent with the literature, while we used $h^{-1}kpc$ throughout the paper, since this is the standard output
of the {\gadget} code, which was used for performing the simulations.

\begin{table*}
\centering
\begin{tabular}{lcccccccccc}
\hline
Nr.		&	satellite	&	$l$ [deg]		&	$b$ [deg]		&	$r_{\odot}$[kpc] 	&	x[kpc	]	&	y[kpc]	&	z[kpc]	&$r_{\text{GC}}$[kpc]	&	$M_{V}$\\
\hline
1	&	LMC			&	280.5	&	-32.9		&	51 		&	-0.50		&	-42.10	&	-27.70	&	50.40	&	-18.1\\
2	&	SMC			&	302.8	&	-44.3		&	64 		&	16.51	&	-38.50	&	-44.70	&	61.26	&	-16.8\\
3	&	Canis Major	&	240.0	&	-0.8		&	7		&	-11.80	&	-6.06		&	-0.098	&	13.27	&	-14.4\\
4	&	Sagittarius	&	5.6		&	-14.2		&	26 		&	16.79	&	2.46		&	-6.378	&	18.12	&	-13.5\\
5	&	Fornax		&	237.1 	&	 -65.7	&	147 		&	-41.16	&	-50.79	&	-133.98	&	149.08	&	-13.4\\
6	&	LeoI			&	226.0    	&	49.1		&	254		&	-123.83	&	-119.63	&	191.99	&	257.88	&	-12.0\\
7	&	Sculptor		&	287.5	&	-83.2		&	86		&	-5.24		&	-9.71		&	-85.40	&	86.11	&	-11.1\\
8	&	LeoII			&	220.2    	&	67.2 		&	233 		&	-77.26	&	-58.28	&	214.79	&	235.59	&	-9.8\\
9	&	Sextans		&	243.5	&	42.3		&	86 		&	-36.68	&	-56.93	&	57.88	&	89.08	&	-9.3\\
10	&	Carina		&	260.1 	&	-22.2		&	105		&	-25.01	&	-95.77	&	-39.67	&	106.64	&	-9.1\\
11	&	Draco		&	86.4		&	34.7		&	76		&	-4.38 	&	62.36	&	43.27	&	76.03	&	-8.8\\
12	&	Ursa Minor	&	105.0	&	44.8		&	76		&	-22.36	&	52.09	&	53.55	&	77.95	&	-8.8\\
13	&	CVn I		&	74.3 		&	79.8		&	218		&	2.15		&	37.16	&	214.56	&	217.76	&	-8.6\\
\hline
14	&	Hercules		&	28.7		&	36.9		&	132 		&	84.29	&	50.69	&	79.27	&	126.32	&	-6.6\\
15	&	Bootes I		&	358.1 	&	69.6		&	66		&	14.69	&	-0.76		&	61.86	&	63.59	&	-6.3\\
16	&	LeoIV		&	265.4	&	57.4		&	154 		&	-14.95	&	-82.70	&	129.74	&	154.58	&	-5.8\\
17	&	Bootes III		&	35.4		&	75.4		&	47		&	1.36		&	6.86		&	45.48	&	46.02	&	-5.8\\
18	&	Ursa Major I	&	159.43	&	54.4		&	97 		&	-61.17	&	19.84	&	78.87	&	101.76	&	-5.5\\
19	&	LeoV		&	261.9	&	58.5		&	178		&	-21.40	&	-92.08	&	151.77	&	178.80	&	-5.2\\
20	&	PscII			&	79.2		&	-47.1		&	182		&	14.91	&	121.70	&	-133.32	&	181.13	&	-5.0\\
21	&	CVn II		&    	113.6	&	 82.7		&	160		&	-16.44	&	18.63	&	158.70	&	160.64	&	-4.9\\
22	&	Ursa Major II	&	152.5	&	37.4		&	32 		& 	-30.85	&	11.74	&	19.44	&	38.30	&	-4.2\\
23	&	Coma Berenices&	241.9	&	83.6		&	44		&	-10.61	&	-4.33		&	43.73	&	45.20	&	-4.1\\
24	&	Bootes II		&	353.7 	&	68.9		&	42		&	6.73		&	-1.66		&	39.18	&	39.79	&	-2.7\\
25	&	Wilkman I		&	158.6	&	56.8		& 	38 		&	-27.67	&	7.59		&	31.80	&	42.83	&	-2.7\\
26	&	Segue II		&	149.4	&	-38.1		&	35		&	-32.01	&	14.02	&	-21.60	&	41.08	&	-2.5\\
27	&	Segue I		&	220.5    	&	50.4		&	23		&	-19.45	&	-9.52		&	17.72	&	27.98	&	-1.5\\
\hline
\end{tabular}
\caption{MW 
satellite galaxies used in this work sorted by the absolute V-band magnitude $M_V$. This list is based on the table given by \citet{McConnachie_2012}, where $l$ and $b$ are angles in
heliocentric galactic coordinates, $r_{\odot}$ is the distance to the sun, $x, y, z$ are cartesian coordinates with the centre of the galaxy in the origin,
and $r_{\text{GC}}$ is the distance to the galactic centre.
}
\label{table:MWsats}
\end{table*}

%***************************************************************%

\end{document}